\documentclass[preprintnumbers, floatfix, preprintnumbers, letterpaper, superscriptaddress,nofootinbib, onecolumn]{revtex4}
\pdfoutput=1
\usepackage{graphicx}
\usepackage{amsmath}
\usepackage{amssymb}
\usepackage{subfigure}
\usepackage{hyperref}
\usepackage{url}
\usepackage{xcolor}
\usepackage{color}
\usepackage{mathrsfs}
\usepackage{calrsfs}
\usepackage{amsfonts}
\usepackage{latexsym}
\usepackage{ragged2e}
\usepackage{textcomp}
\usepackage{phaistos}
\usepackage{epstopdf}
\makeatletter
\renewcommand\@makefnmark{\hbox{\@textsuperscript{\normalfont\color{purple}\@thefnmark}}}
\renewcommand\@makefntext[1]{%
  \parindent 1em\noindent
            \hb@xt@1.8em{%
                \hss\@textsuperscript{\normalfont\@thefnmark}}#1}
\makeatother
\usepackage{caption}
\DeclareCaptionJustification{justified}{\leftskip=0pt \rightskip=0pt \parfillskip=0pt plus 1fil}
\definecolor{vividviolet}{rgb}{0.62, 0.0, 1.0}
\definecolor{amaranth}{rgb}{0.9, 0.17, 0.31}
\definecolor{palatinateblue}{rgb}{0.15, 0.23, 0.89}
\definecolor{brightpink}{rgb}{1.0, 0.0, 0.5}
\definecolor{cornflowerblue}{rgb}{0.39, 0.58, 0.93}
\definecolor{deepcarminepink}{rgb}{0.94, 0.19, 0.22}
\definecolor{radicalred}{rgb}{1.0, 0.21, 0.37}

\hypersetup{ linktoc=all,
    colorlinks, linkcolor={palatinateblue},
    citecolor={brightpink}, urlcolor={amaranth}
}

\graphicspath{{Images/}}

\renewcommand{\d}[1]{\ensuremath{\operatorname{d}\!{#1}}}

\graphicspath{{Images/}}
\renewcommand{\d}[1]{\ensuremath{\operatorname{d}\!{#1}}}
\makeatletter
\def\@fnsymbol#1{\ensuremath{\ifcase#1\or $\PHplaneTree$ \or $\textleaf$ 
\else\@ctrerr\fi}}%

\makeatother

\def\sideremark#1{\ifvmode\leavevmode\fi\vadjust{\vbox to0pt{\vss
 \hbox to 0pt{\hskip\hsize\hskip1em
 \vbox{\hsize1.5cm\tiny\raggedright\pretolerance10000
 \noindent #1\hfill}\hss}\vbox to8pt{\vfil}\vss}}}%
\def\sideremark#1{\ifvmode\leavevmode\fi\vadjust{\vbox to0pt{\vss
 \hbox to 0pt{\hskip\hsize\hskip1em
 \vbox{\hsize1.3cm\tiny\raggedright\pretolerance10000
 \noindent #1\hfill}\hss}\vbox to8pt{\vfil}\vss}}}%
\begin{document}

\title{The Interior Volume of Kerr-AdS Black Holes}

\author{Xiao Yan \surname{Chew}}
\email{xychew998@gmail.com}
\affiliation{Center for Gravitation and Cosmology, College of Physical Science and Technology, Yangzhou University, \\180 Siwangting Road, Yangzhou City, Jiangsu Province  225002, China}
\affiliation{Department of Physics Education, Pusan National University, Busan 46241, Republic of Korea}
\affiliation{Research Center for Dielectric and Advanced Matter Physics, Pusan National University, Busan 46241, Republic of Korea}

\author{Yen Chin \surname{Ong}}
\email{ycong@yzu.edu.cn}
\affiliation{Center for Gravitation and Cosmology, College of Physical Science and Technology, Yangzhou University, \\180 Siwangting Road, Yangzhou City, Jiangsu Province  225002, China}
\affiliation{School of Aeronautics and Astronautics, Shanghai Jiao Tong University, Shanghai 200240, China}

\begin{abstract}
The interior volume of black holes as defined by Christodoulou and Rovelli exhibits many surprising features. For example, it increases with time, even under Hawking evaporation. For some black holes, the interior volume is not even a monotonic increasing function of its area, which means one cannot infer how large a black hole is by just looking from the outside. Such a notion of volume, however, turns out to be useful in the context of holography, as it seems to be dual to the complexity of the boundary field theory. In this study, we investigate the properties of the interior volume of 4-dimensional Kerr-AdS black holes, fixing either the mass parameter $M$ or the physical mass $E$, whilst varying the values of the cosmological constant. We found that the volume as a function of the radial coordinate features a ``double lobe'' while fixing $M$, whereas fixing $E$ yields behaviors that are qualitatively similar to the asymptotically flat case. We briefly comment on the holographic complexity of Kerr-AdS black holes.
\begin{center}

\end{center}
\end{abstract}

\maketitle

\section{Introduction: Black Hole Interior Volume}

Black holes are remarkable objects that continue to surprise us even after a century of general relativity. The defining feature of a black hole is that nothing that enters its event horizon can escape from its interior. Therefore an exterior observer remains ignorant about just how the interior spacetime looks like. In particular, how big is a black hole from the inside? Take a spherical surface of area $4\pi R^2$ in Euclidean space $\Bbb{R}^3$, it bounds a space of volume $4\pi R^3/3$. However, this is not the case for black holes, which, after all, has nontrivial spacetime curvature, which allows a fixed area to bound a larger than expected volume. Indeed, the first challenge is to define just what we mean by (spatial) ``volume'' (see \cite{0801.1734} for some options). Unlike the event horizon, which is uniquely defined in Lorentzian geometry, there is a plethora of ways to define volume of a black hole. In this work our suject of study is the Christodoulou-Rovelli volume \cite{Christodoulou:2014yia}. Hereinafter, by volume, we always mean the Christodoulou-Rovelli volume. 

The idea of Christodoulou and Rovelli is to look for the volume of the \emph{largest} spacelike spherically symmetric surface bounded by the event horizon of the black hole, which is a geometrical property that is coordinate independent. 
For a 4-dimensional asymptotically flat Schwarzschild black hole, the volume is 
\begin{equation}
\text{Vol.} = \int^{v} \int_{S^2} \text{max} \left(r^2  \sqrt{\frac{2M}{r}-1}\right) \sin \theta~ \d \theta \d \phi \d v',
\end{equation}
 in which $v$ is the usual advanced time in the Eddington--Finkelstein coordinates. (Note that $v$ is a spacelike direction while $r$ is a timelike one inside the black hole.)
The lower limit of the integral is ignored because it only gives sub-leading, negligible, contribution to the total volume at late time. The function $r^2\sqrt{2M/r-1}$ is maximized at $r=3M/2$, which gives, when $v$ is large,
\begin{equation}
\text{Vol.}=3\sqrt{3}\pi M^2 v.
\end{equation}

We see that the interior volume grows linearly in time $v$. This result is entirely classical and is in no way controversial. If we accept that black holes are solutions to the Einstein field equations, then we should accept that the volume defined in this manner inside a black hole continues to grow linearly with time, although its area stays fixed (the volume continues to increase even if the area is decreasing under Hawking evaporation \cite{1503.08245, 1604.07222}). This is not so surprising if we recall that the interior spacetime of a static solution like Schwarzschild is nevertheless \emph{not} static. 

Assuming the Schwarzschild geometry, Christodoulou and Rovelli estimated that Sagittarius $\text{A}^*$, the supermassive black hole at the center of our Milky Way, contains enough volume to fit a \emph{million} solar systems. Its areal radius, on the other hand, is only about 10 times larger than the Earth-Moon distance \cite{Christodoulou:2014yia}. Generalization of the Christodoulou-Rovelli volume to the case of asymptotically flat Kerr black hole does not change this result by much \cite{Bengtsson:2015zda}, despite the rotation rate of Sagittarius $\text{A}^*$ is about $90\%$ of the extremal limit.

Generalizations to topological black holes in locally asymptotically anti-de Sitter (AdS) spacetimes further reveal surprising features of the interior volume: for some black holes, the volume is not a monotonic function of the area, which means that one cannot judge from the outside how large the volume is even if the time $v$ is known \cite{1503.01092}. In fact, the AdS generalizations are important, for there is evidence that the interior volume is holographically dual to the (quantum) complexity of the boundary field theory in the context of gauge/gravity correspondence \cite{1402.5674, 1403.5695, 1406.2678, 1411.0690, 1509.06614}, which also grows linearly in time (Christodoulou-Rovelli volume can be defined for maximally extended ``two-sided'' black holes as well) at least for a time proportional to the exponential of the black hole entropy \cite{1507.02287}. This is now referred to as the ``Complexity-Volume'' (CV) Conjecture. Another proposal is that complexity is dual to the action of a certain spacetime region referred to as the ``Wheeler-DeWitt patch''; this is known as the ``Complexity-Action'' (CA) Conjecture \cite{1509.07876, 1512.04993}. Subsequently, many properties of both interior volumes and Wheeler-DeWitt action had been studied, the similarities and differences between CV and CA Conjectures investigated (see, e.g., \cite{1606.08307, 1612.00433, 1709.10184, 1712.09706v2, Couch:2018phr, Yang:2019alh, 1804.01561, 1811.10650, 1902.02475}).

In this work, we study the interior volume of 4-dimensional Kerr-AdS black holes. We shall fix the mass of the black hole (either the physical mass $E$ or the ``geometric mass parameter'' $M$ -- see next section). In view of the recent interests in ``black hole chemistry'' \cite{1608.06147} -- in which the negative cosmological constant in AdS bulk is treated as thermodynamical pressure \cite{0904.2765,1106.6260, 1209.1272}, and holographically dual to the number of colors ($N_c$) of the boundary field theory via the relation $|\Lambda| \sim N_c^{-4/3}$ \cite{1409.3575} -- we consider the effect of varying cosmological constant on the geometry of the interior volume. 
We find that fixing the physical mass gives rise to a ``double lobe'' feature in the plot of the volume as function of coordinate radius, which is absent if we fix instead the geometric mass.
Our work is purely focusing on the black hole properties, though we hope that our study might be helpful for holography in the future. We will work in the Planck units so that $\hbar=c=G=k_B=1$. In the appendix, we attach two plots of the interior volume of Kerr--AdS black hole with mass $M$ and the physical mass $E$ in the unit of $L$.

\section{Some Properties of Kerr--AdS Black Hole}

The metric of Kerr--AdS black hole in Boyer--Lindquist coordinates $(t,r,\theta,\phi)$ in 4-dimensions is given by \cite{McInnes:2015vga}
\begin{equation}
\d s^2 =-  \frac{\Delta_r}{\Sigma} \left( \d t- \frac{a}{\Xi} \sin^2 \theta \d\phi \right)^2  +  \Sigma \left( \frac{\d r^2}{\Delta_r} + \frac{\d \theta^2}{\Delta_\theta} \right) + \frac{ \Delta_\theta \sin^2 \theta}{\Sigma}  \left(a \d t- \frac{\Sigma}{\Xi}  \d\phi \right)^2 \,, 
\end{equation}
where
\begin{equation}
\Sigma :=  r^2 +a^2  \cos^2 \theta \,, \quad  \Delta_r := (r^2+a^2) \left( 1-\frac{\Lambda r^2}{3}  \right)-2 M r \,, \quad  
\Delta_\theta := 1+\frac{a^2 \Lambda}{3} \cos^2 \theta \,, \quad
\Xi := 1+\frac{a^2 \Lambda}{3} \,.
\end{equation}

The cosmological constant $\Lambda$ is related to $L$, the asymptotic curvature radius of anti--de--Sitter spacetime in 4-dimensions, via $\Lambda = - 3/L^2$. Note that the metric has $\Xi$ in the denominator, which puts a bound on the rotation parameter: $a < L$. This bound is distinct from the cosmic censorship bound. See \cite{1108.6234} for more discussions\footnote{The case $a>L$ for AdS-Kerr black hole in 5-dimensions is discussed in \cite{1906.01169, 1911.08222}. Since the geometry for $a>L$ is quite different from the $a<L$ case (for example, there are closed causal curves for sufficiently small $\theta$), and furthermore starting with $a<L$, one cannot cross the $a=L$ divide to reach $a>L$ branch of the solution, we shall restrict our investigation to the $a<L$ case.}.

Note that the parameter $M$ is not the physical mass of Kerr--AdS. 
The physical mass $E$ of Kerr--AdS is related to $M$ by \cite{0408217}
\begin{equation}
E = \frac{M}{\Xi^2} \,.
\end{equation}
The physical mass, also known as the Abbott-Deser mass \cite{AD}, is the conserved quantity that corresponds to the timelike Killing vector of the geometry. 
Naturally it plays crucial roles in physical processes. Crucially, it is this mass that enters the second law of thermodynamics of black holes; working with $M$ instead of $E$ might give misleading results that the second law could be violated under physical processes \cite{McInnes:2015vga}. 

On the other hand, the mass parameter $M$ is not without advantages. It is a ``geometric parameter'' that gives better intuiton regarding the underlying spacetime geometry of the black hole. To illustrate this, an example was given in \cite{McInnes:2015vga}: Consider an AdS-Kerr black hole with a very large physical mass, say $E=2 \times 20^6$ (in the units such that $L=1$), and large angular momentum such that $a^2 = 0.99$. The black hole horizon is located at $r \approx 1.0004998$. One might expect that such a large mass concentrated at such a small value of the radial coordinate would distort the geometry by a large amount. However it turns out that the Gaussian curvature of the horizon is only of order unity (as function of $\theta$, it ranges between $-\pi/2$ and $\pi/2$). However, the mass parameter is only $M=2$ in this example. Thus $M$ can give a better indication of the geometry than the physical mass (although we are not being very precise here). In other words, depending on one's purpose, $E$ and $M$ can both be useful parameters.

\begin{figure} [h!]
\begin{center}
\mbox{(a)
\includegraphics[angle=270,scale=0.31]{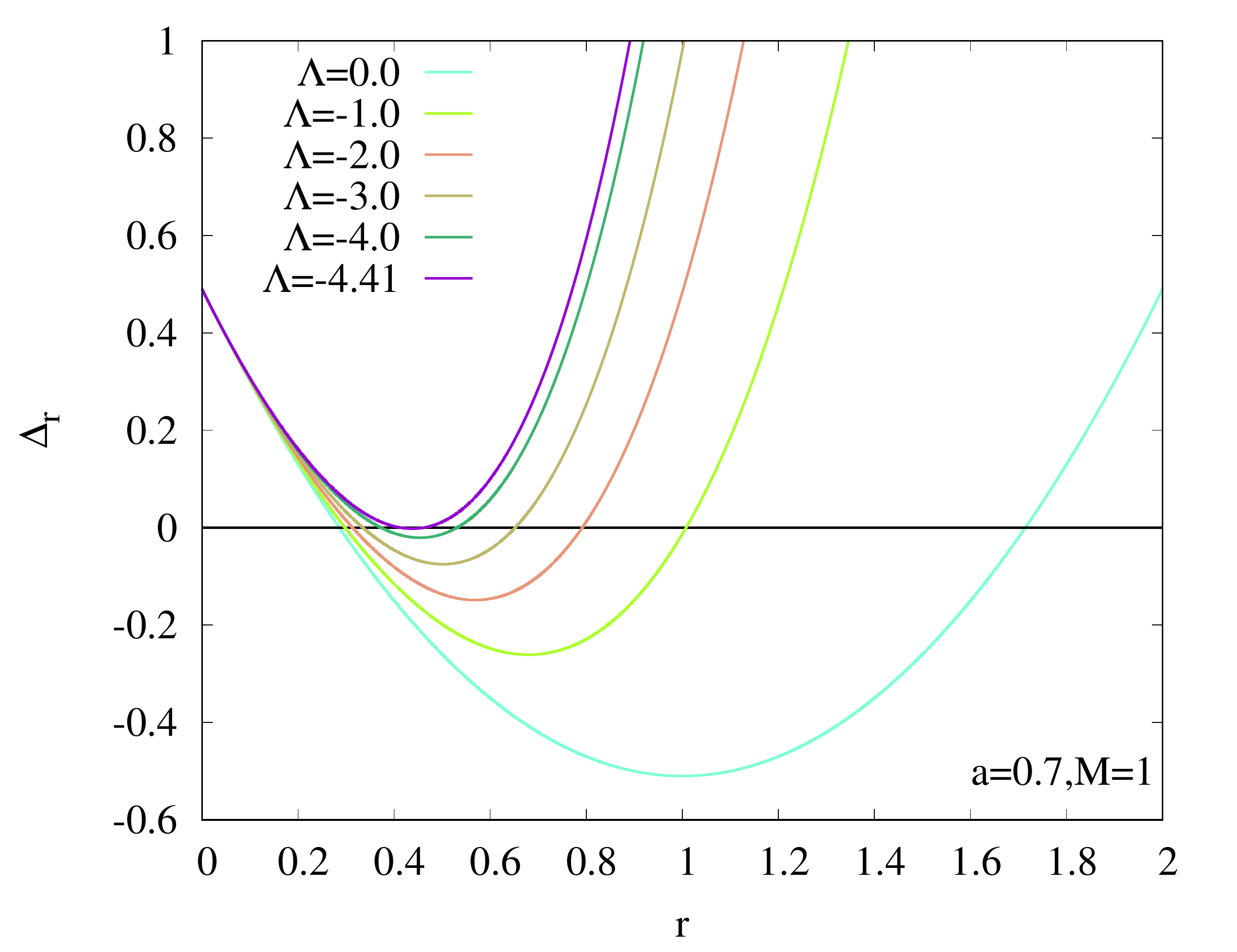}
(b) 
\includegraphics[angle=270,scale=0.31]{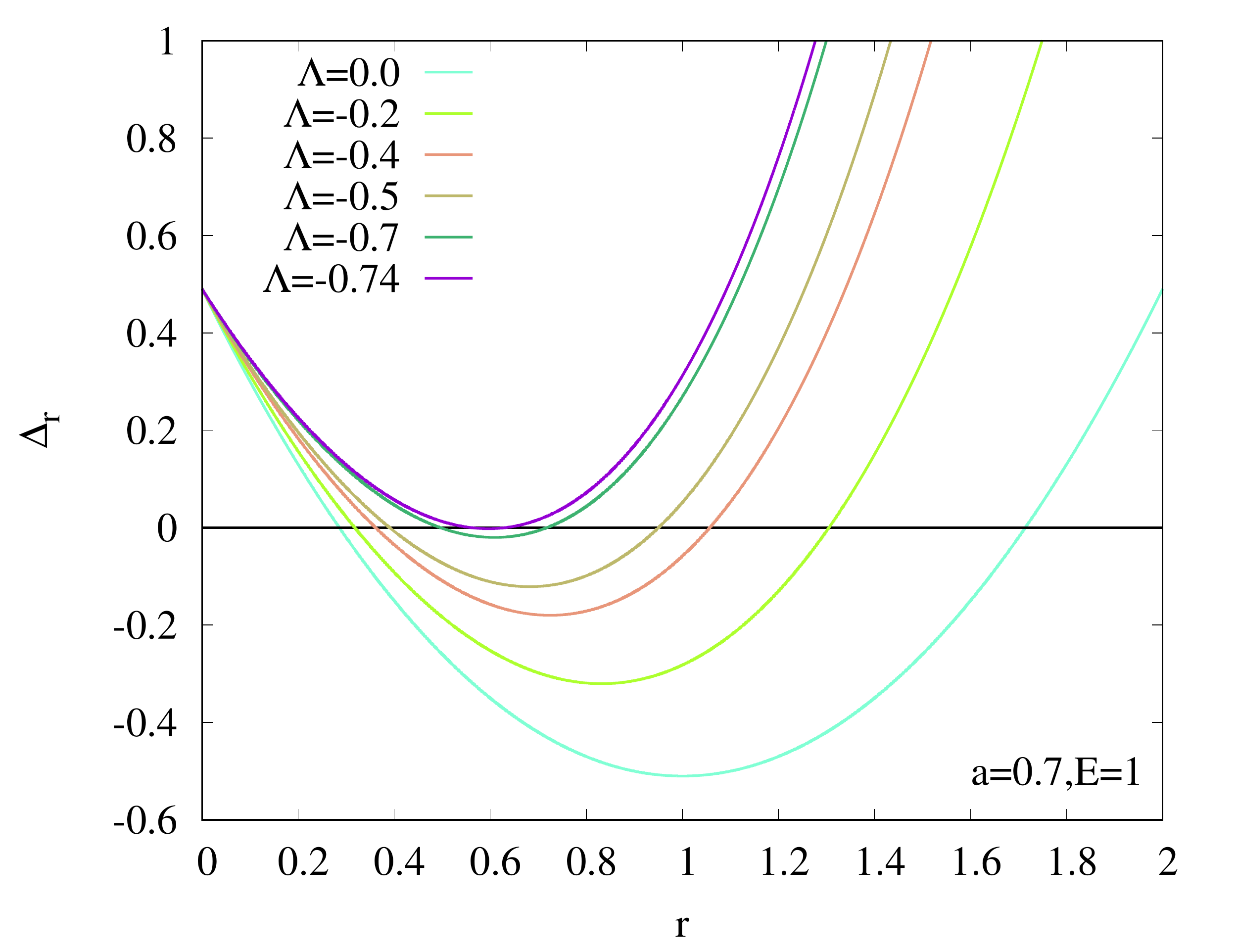} 
}
\end{center}
\caption{The plots for the function $\Delta_r$ with $a=0.7$ for (a) $M$=1 and (b) $E$=1. \label{Fig1}}
\end{figure}

The horizons of Kerr--AdS black holes can be obtained by solving $\Delta_r=0$. Since the function $\Delta_r$ is a fourth order polynomial, it has either 4 real roots or 2 real roots in principle. Here we exhibit the behaviour of $\Delta_r$ in Fig. \ref{Fig1} with $a=0.7$ by fixing $M=1$ and $E=1$, respectively. The values are chosen for simplicity only. Indeed, the qualitative features of our results remain the same if we change the values of $M$ or $E$.

When $M=1$, the function $\Delta_r$ only has 2 real roots (in fact this is true for all  Kerr-AdS black holes). The largest root represents the outer horizon, $r_{+}$ while the smallest root represents the inner (Cauchy) horizon, $r_{-}$. Of course, if we fixed $E=1$ instead, the Kerr--AdS black hole also has outer and inner horizons, which is similar to the case with $M=1$. When $a=0$, the Kerr--AdS black hole reduces to Schwarzschild--AdS black hole which only has a single horizon. Furthermore, Kerr--AdS black hole also possesses a coordinate singularity on the rotation axis which is determined from the condition $\Delta_\theta=0$.

The extremality of Kerr--AdS black hole can be determined by the condition \cite{Agnese:1999df}
\begin{equation} \label{extreme_cond}
 \Delta_r = \frac{\text{d}}{\text{d} r} \Delta_r = 0 \,.
\end{equation}
When the above condition is satisfied, the outer and inner horizons of Kerr--AdS black hole coincide, and its Hawking temperature vanishes.

\subsection{The Domain of Existence}



Kerr--AdS black holes are determined by 3 parameters, namely the cosmological constant $\Lambda$, the rotation parameter $a$ and the mass parameter (either $M$ or $E$). For convenience, we fix the mass of black hole by setting either $M=1$ or $E=1$\footnote{Equation \eqref{extreme_cond} and the volume \eqref{Volume_eqn} have the following scaling symmetry: $r \rightarrow \kappa r$, $a \rightarrow \kappa a$, $\Lambda \rightarrow \kappa^{-2}\Lambda$ and $M \rightarrow \kappa M$. Since $\Lambda=-3/L^2$, we see that $L$ can be scaled as $L \rightarrow \kappa L$. In our case, we fix the value of $\Lambda$ and scale the remaining parameters by $M$, thus the parameters $r$, $a$ and $\Lambda$ are in the unit of mass. Furthemore, if we fix the mass and scale the ramaining parameters by $L$, then $r$, $a$ and $M$ are in the unit of $L$.}

\begin{figure}[h!] 
\centering
\includegraphics[angle=270,scale=0.31]{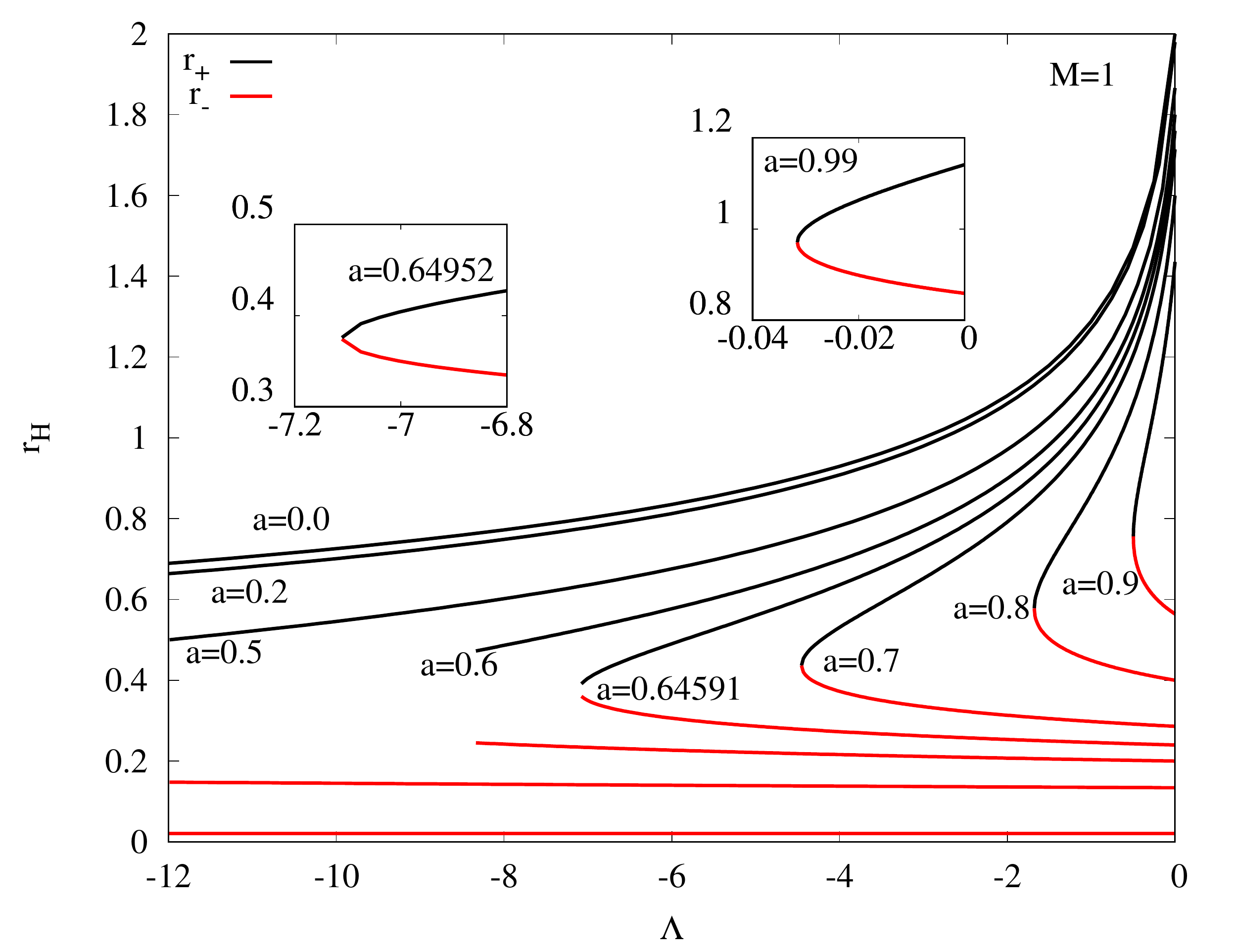} 
\caption{The domain of existence for Kerr--AdS black hole with $M=1$ for $a < L$. The outer horizon is represented by the black curves and the inner horizon is represented by the red curves. The two curves meet when black hole is extremal. The curves will be gapless for any $a \gtrapprox 0.64952$, but otherwise features a gap between the outer and inner horizon.\label{Fig2}}
\end{figure}

\begin{figure} [t!]
\begin{center}
\mbox{(a)
\includegraphics[angle=270,scale=0.31]{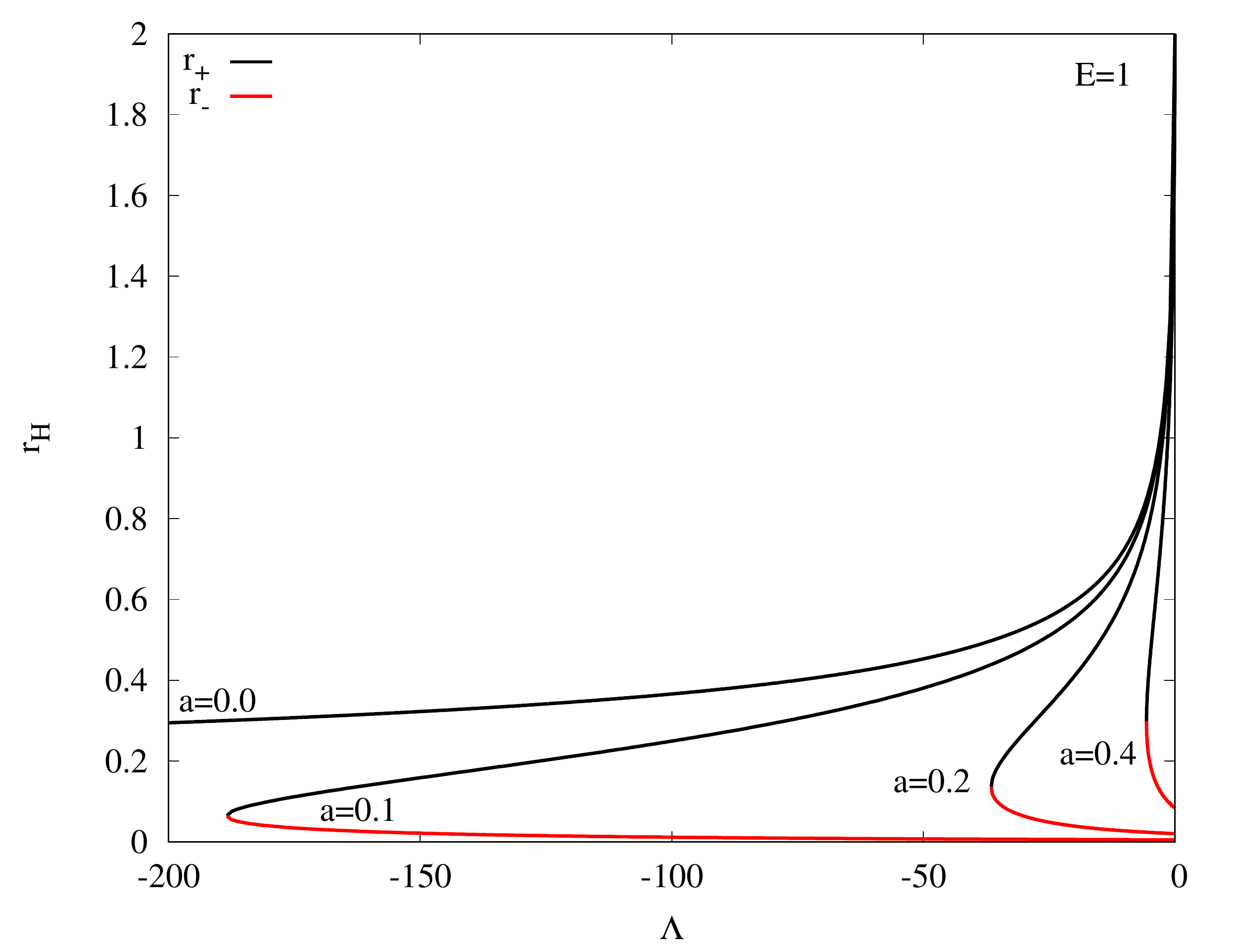}
(b) 
\includegraphics[angle=270,scale=0.31]{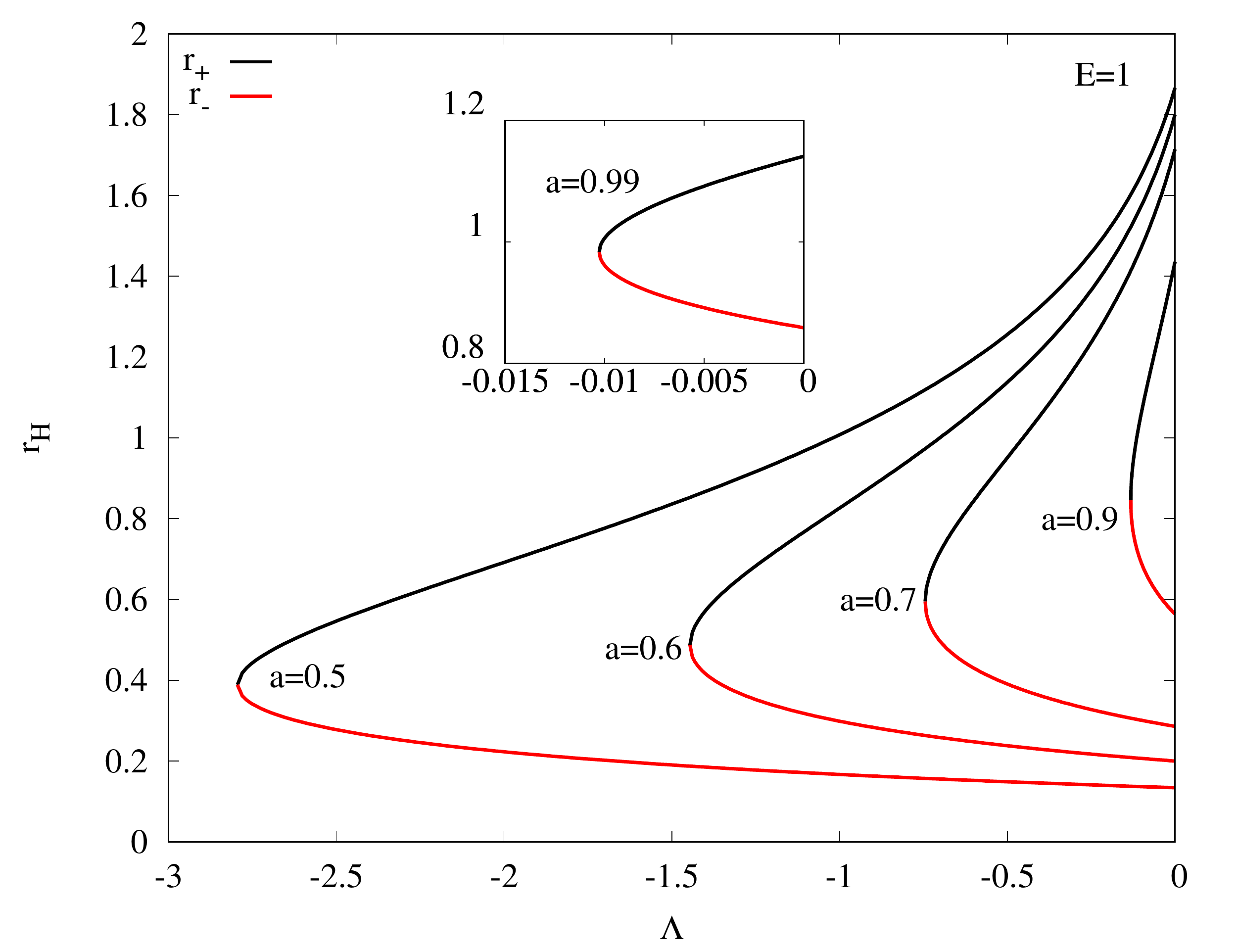} 
}
\end{center}
\caption{The domain of existence for Kerr--AdS black holes for several values of $a$ with $E=1$. The black curves denote the outer horizon and the red curves denote the inner horizon. The two curves meet at a particular value of the cosmological constant, $\Lambda_{\text{maximal}}$, which corresponds to the extremal Kerr--AdS black hole. The curves are all gapless, c.f. Fig.(\ref{Fig2}). \label{Fig3}}
\end{figure}

We then study the domain of existence for Kerr--AdS black holes by fixing the value of $a$ and varying $\Lambda$ to calculate the outer and inner horizon of Kerr--AdS black hole. We exhibit the domain of existence of Kerr--AdS black hole for $M=1$ in Figure \ref{Fig2} and for $E=1$ in Figure \ref{Fig3}. 

In Figure \ref{Fig2}, the outer horizon attains its maximum value when $a=0$, i.e., when the black hole is the static AdS-Schwarzschild black hole. When we decrease $\Lambda$ from zero with $a$ stays fixed, the coordinate value of the outer horizon decreases but the inner horizon increases, and at the end both horizons meet, and the black hole becomes extremal. We also observe that all else being equal, Kerr-AdS black holes become extremal at smaller value of $|\Lambda|$ for larger value of the rotation parameter $a$. Note that the bound $a<L$ translates to $a < \sqrt{-3/\Lambda}$. That is, with $|\Lambda|$ large, the allowed values of $a$ becomes smaller. 
This explains the gap in the plot between the outer and inner horizons. Indeed, for fixed $M=1$, the curve will be gapless for any $a \gtrapprox 0.64952$. (This value of course changes if $M = \text{const.} \neq 1$.)

In Figure \ref{Fig3}, analogous to the case $M=1$, the Kerr--AdS black holes with $E=1$ also show the same qualitative behaviour, that is the outer horizon increases but the inner horizon decreases, and then both horizons meet to form extremal black hole. Again, the black holes attain extremality at smaller value of $|\Lambda|$ for larger value of $a$. However, it turns out the curves are always gapless. We emphasize that the properties discussed in this subsection hold also for other values of $M$ and $E$.

\subsection{Thermodynamical Properties}

Although our result is classical, for the sake of completeness we shall give a short summary of the entropy and temperature of AdS-Kerr black holes.
The Bekenstein-Hawking entropy of a Kerr--AdS black hole is given by the usual formula
\begin{equation}
S\equiv S_H=\frac{A_H}{4}, 
\end{equation}
in which $A_H$ is the area of the event horizon, given by \cite{McInnes:2015vga}
\begin{equation}
A_H = \frac{4 \pi (r_{+}^2+a^2)}{\Xi} \,.
\end{equation}
The Hawking temperature is given by
\begin{equation}
T_H = \frac{r_{+} \left( 1 - \frac{a^2 \Lambda}{3} -\Lambda r_{+}^2 - \frac{a^2}{r_{+}^2} \right)}{4 \pi (r_{+}^2+a^2) } \,.
\end{equation}

\begin{figure}[t!]
\centering
\mbox{ 
(a)  \includegraphics[angle =270,scale=0.31]{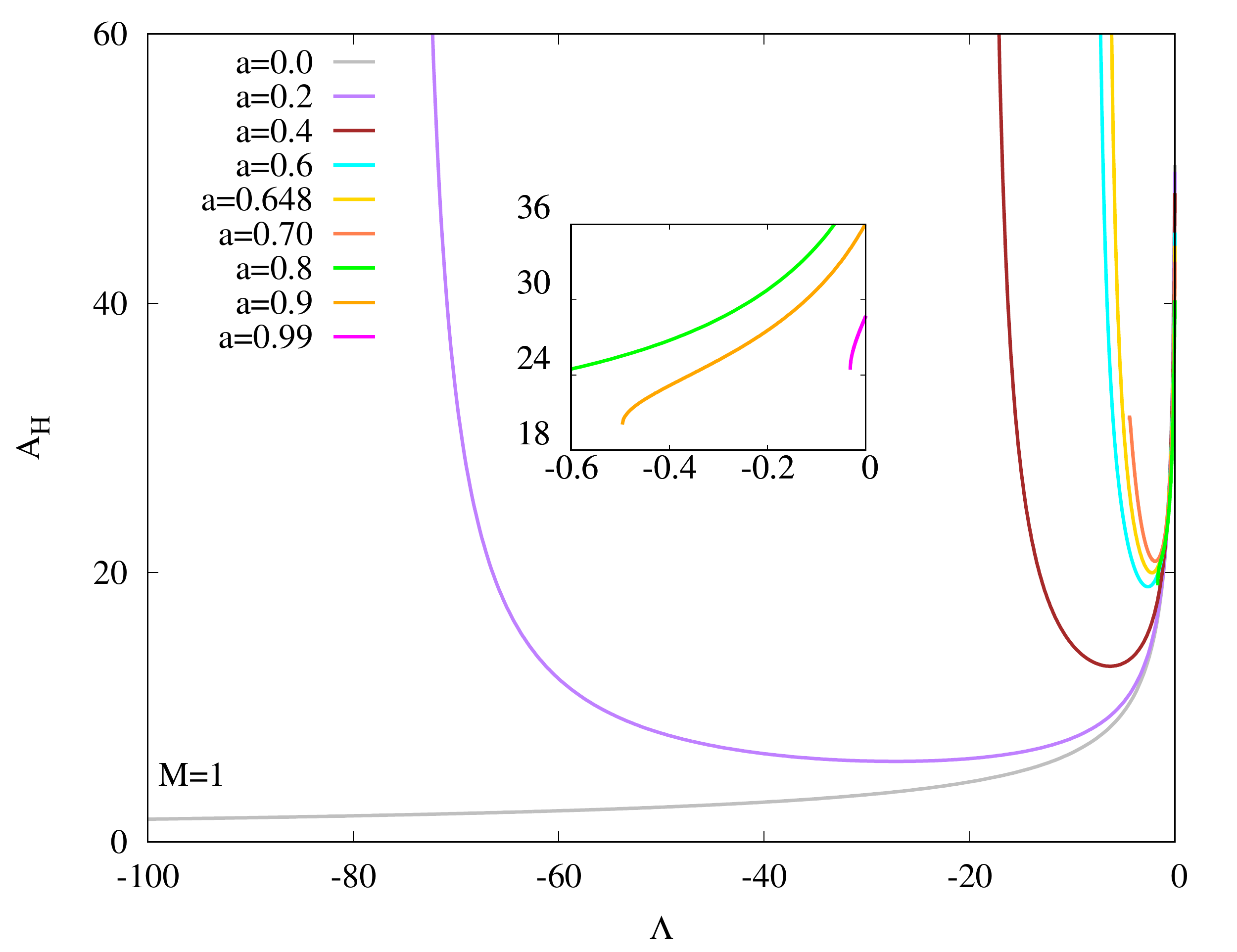}
(b) \includegraphics[angle =270,scale=0.31]{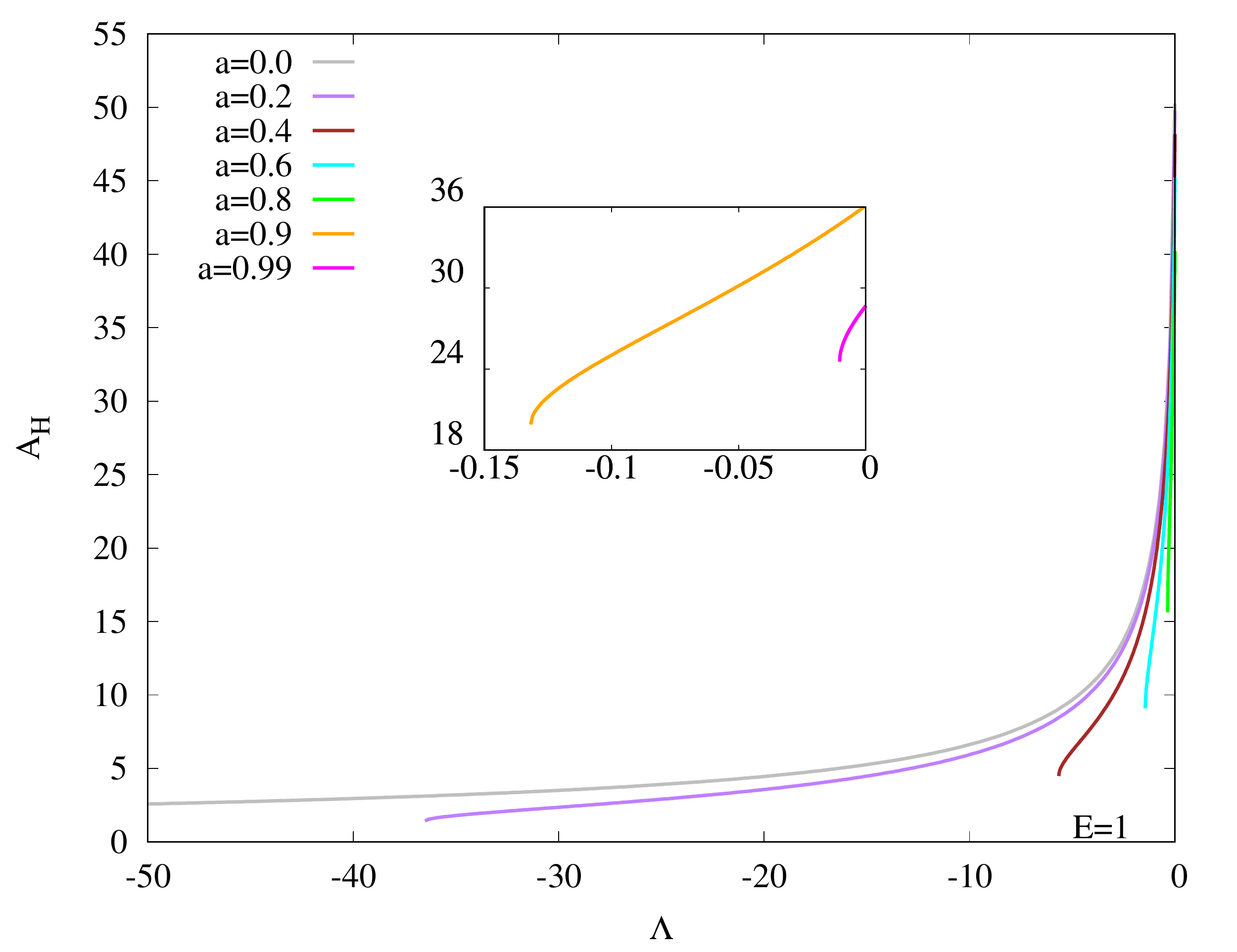} 
}
\mbox{
(c)
\includegraphics[angle =270,scale=0.31]{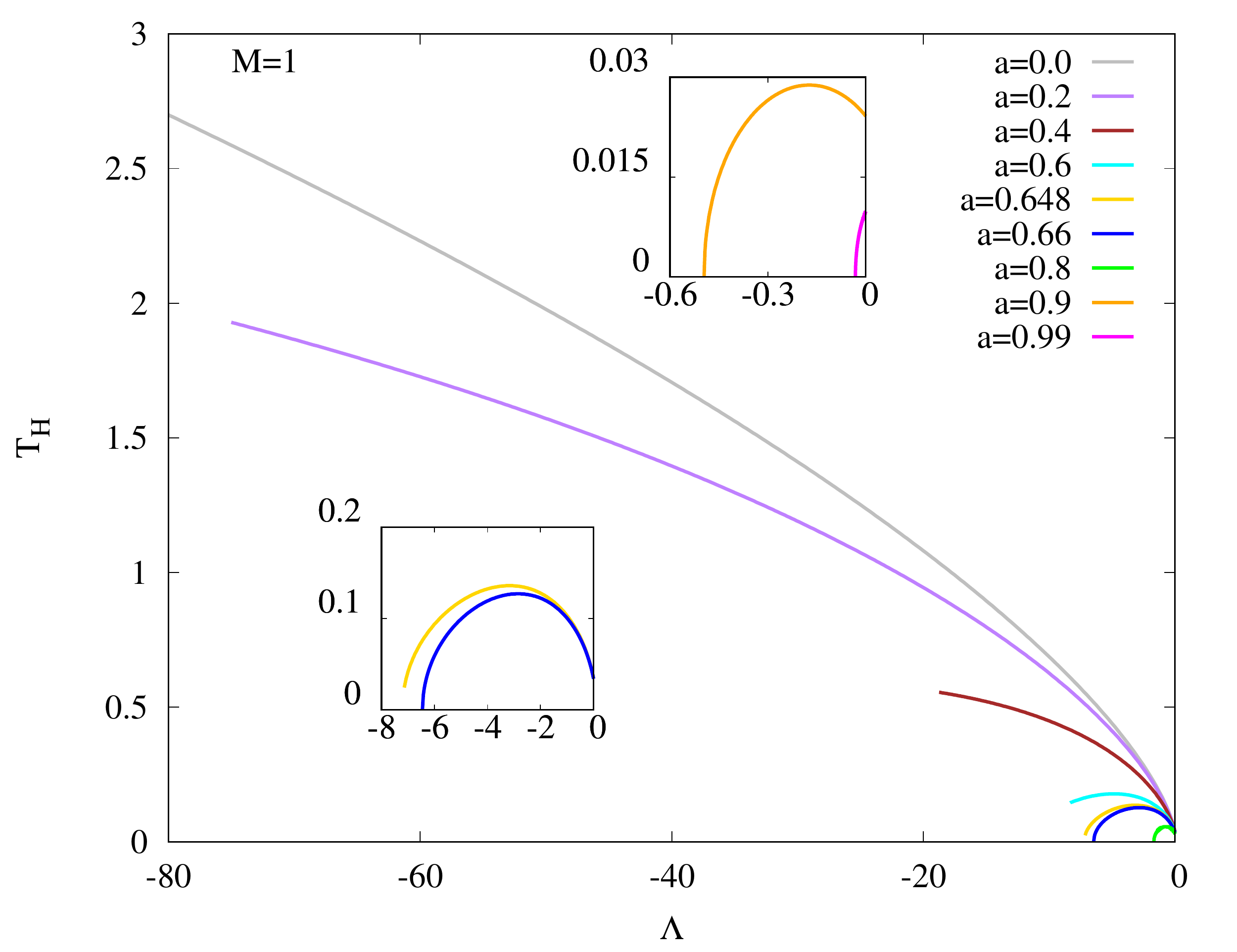}
(d)
\includegraphics[angle =270,scale=0.31]{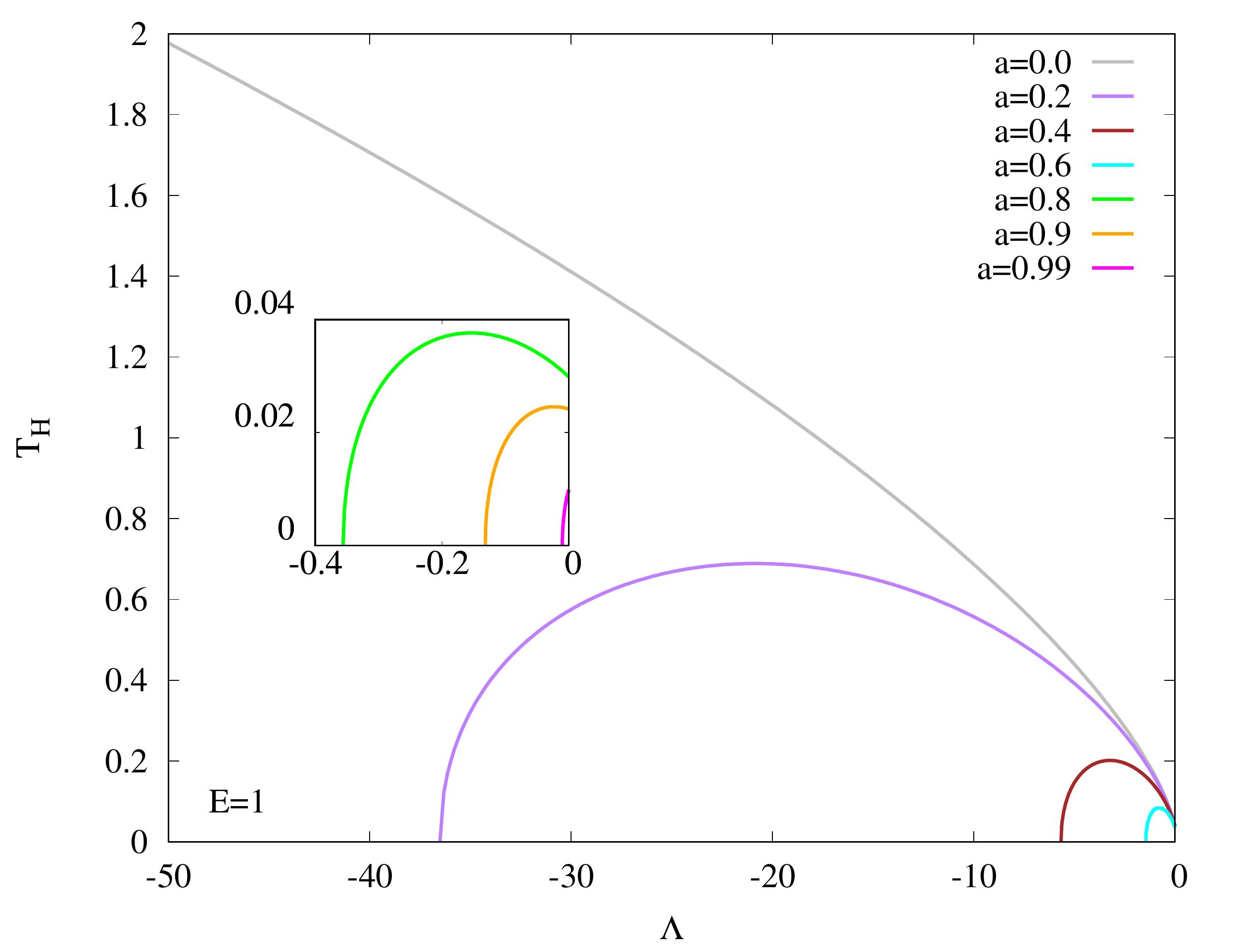}
}
\caption{The area of horizon $A_H$ of Kerr--AdS black holes for several values of $a$ with (a) $M=1$ and (b) $E=1$; The Hawking temperature $T_H$ of Kerr--AdS black holes for various values of $a$ for (c) $M=1$ and (d) $E=1$. \label{Fig4}}
\end{figure}

Figures  \ref{Fig4} (a) and (b) 
show that the area of horizon decreases when $\Lambda$ decreases from 0 downwards, at least initially. The area of horizon $A_H$ for Kerr--AdS black holes with $M=1$ steadily decreases to a minimum value for  $a \gtrapprox 0.64952$. However, $A_H$ turns around and eventually diverges for $a \lessapprox 0.64952$ when the black holes approaches the bound $a<L$. This turn-over behavior is not present in the case $E=1$, the area of the black hole simply decreases to a minimal value when $\Lambda$ decreases. Eventually the black hole becomes extremal.

Figures \ref{Fig4}  (c) and 4 (d) show that as $\Lambda$ decreases, the Hawking temperature $T_H$ of Kerr--AdS black holes with $E=1$ increases to a maximum value, and then decreases to zero, corresponding to an extremal black hole state. Note that the maximum of $T_H$ shifts to a larger value of $\Lambda$ (i.e. smaller $|\Lambda|$) when $a$ increases. The Kerr--AdS black holes with $M=1$ also exhibit the similar qualitative behaviour for the Hawking temperature, although for $a \lessapprox 0.64952$ the extremal state is never reached.

\section{Interior Volume of AdS-Kerr Black Holes}

Here, we follow the approach of \cite{Bengtsson:2015zda} to calculate the interior volume of Kerr--AdS black holes. In order to calculate the interior volume of Kerr--AdS, we first transform the Boyer--Lindquist coordinates into the Eddington--Finkelstein coordinates $(\nu,r,\theta,\tilde{\phi})$ by introducing the following coordinate transformation:
\begin{equation}
\nu = t + H(r)  \,,  \qquad \qquad \tilde{\phi} = \phi + G(r) \,,
\end{equation}
where $H(r)$ and $G(r)$ are arbitrary real--valued functions. We can obtain the expressions of $H(r)$ and $G(r)$ by substituting the differential 
\begin{equation} 
\text{d}\nu = \text{d} t + H'(r)\,\text{d} r  \,,  \qquad \qquad \text{d}\tilde{\phi} = \text{d}\phi + G'(r)\,\text{d} r \,,
\end{equation}
into the metric of Kerr--AdS in Boyer--Lindquist form, with the condition that $g_{rr}=0$ and $g_{r \nu}=2$ being imposed. This yields the following expressions for $H(r)$ and $G(r)$:
\begin{equation}
H(r) = \frac{r^2+a^2}{\Delta_r} \,, \qquad  G(r) = \frac{a \Xi}{\Delta_r}  \,.
\end{equation}


Therefore, the metric of Kerr--AdS spacetime in Eddington--Finkelstein coordinates is
\begin{align}
\text{d} s^2  &= -  \frac{\Delta_r - a^2 \sin^2 \theta \Delta_\theta }{\Sigma} \text{d} \nu^2 + 2 \text{d} r \text{d} \nu + \frac{\Sigma}{\Delta_\theta} \d\theta^2 -\frac{2 a \sin^2 \theta }{\Xi} \text{d} r \text{d} \tilde{\phi}  - \frac{2 a \sin^2 \theta \, \left[ (r^2+a^2) \Delta_\theta- \Delta_r \right] }{\Sigma \Xi} \text{d} \nu\text{d} \tilde{\phi}  \nonumber \\
 & \qquad -   \frac{ \sin^2 \theta \left[ a^2 \sin^2 \Delta_r-  (r^2+a^2)^2 \Delta_\theta \right] }{\Sigma \Xi^2}  \text{d}\tilde{\phi}^2 \,.
\end{align}

The interior volume with constant hypersurface $r$ for Kerr--AdS black hole can now be calculated as follows\footnote{Although this is a direct generalization of the Christodoulou-Rovelli volume to the rotating case, as emphasized in \cite{Bengtsson:2015zda} there is a difference. In fact, the hypersurface $r=3M/2$  that maximizes the volume of an asymptotically flat Schwarzschild black hole has a special property: it is the solution of the polynomial $f(M; r)=0$ obtained by setting the trace of the second fundamental form to zero \cite{bruce}. For asymptotically flat Kerr black hole, the vanishing trace condition yields $f(M, a, \theta; r)=0$, which is a polynomial in $r$ of degree 7. There does not exist $r=r(M,a)$ such that $f$ vanishes. So in this sense, what we computed here is not strictly the maximal volume, though it is arguably close to one \cite{Bengtsson:2015zda}. Again, defining volume is a tricky business in general, see \cite{1703.02396} for a recent rigorous attempt.}:
\begin{align}
\text{Vol.}(r) &= \int^{v}  \sin \theta \frac{ \sqrt{ |\Delta_r| \Sigma}}{\Xi}~ \d v \d \theta \text{d} \tilde{\phi} \\
&= \frac{\pi \sqrt{ |\Delta_r|}}{a \Xi}  \left[ 2 a \sqrt{r^2+a^2 } + r^2  \ln \left(  \frac{\sqrt{r^2+a^2 } +a }{\sqrt{r^2+a^2 } -a}  \right)  \right] v\,.  \label{Volume_eqn}
\end{align} 
\emph{Henceforth we ignore $v$}, since $v$ only governs the linear growth of the volume. We are interested in the coefficient, which are nontrivial.

\begin{figure}[h!]
\centering
\mbox{ 
(a)  \includegraphics[angle =270,scale=0.31]{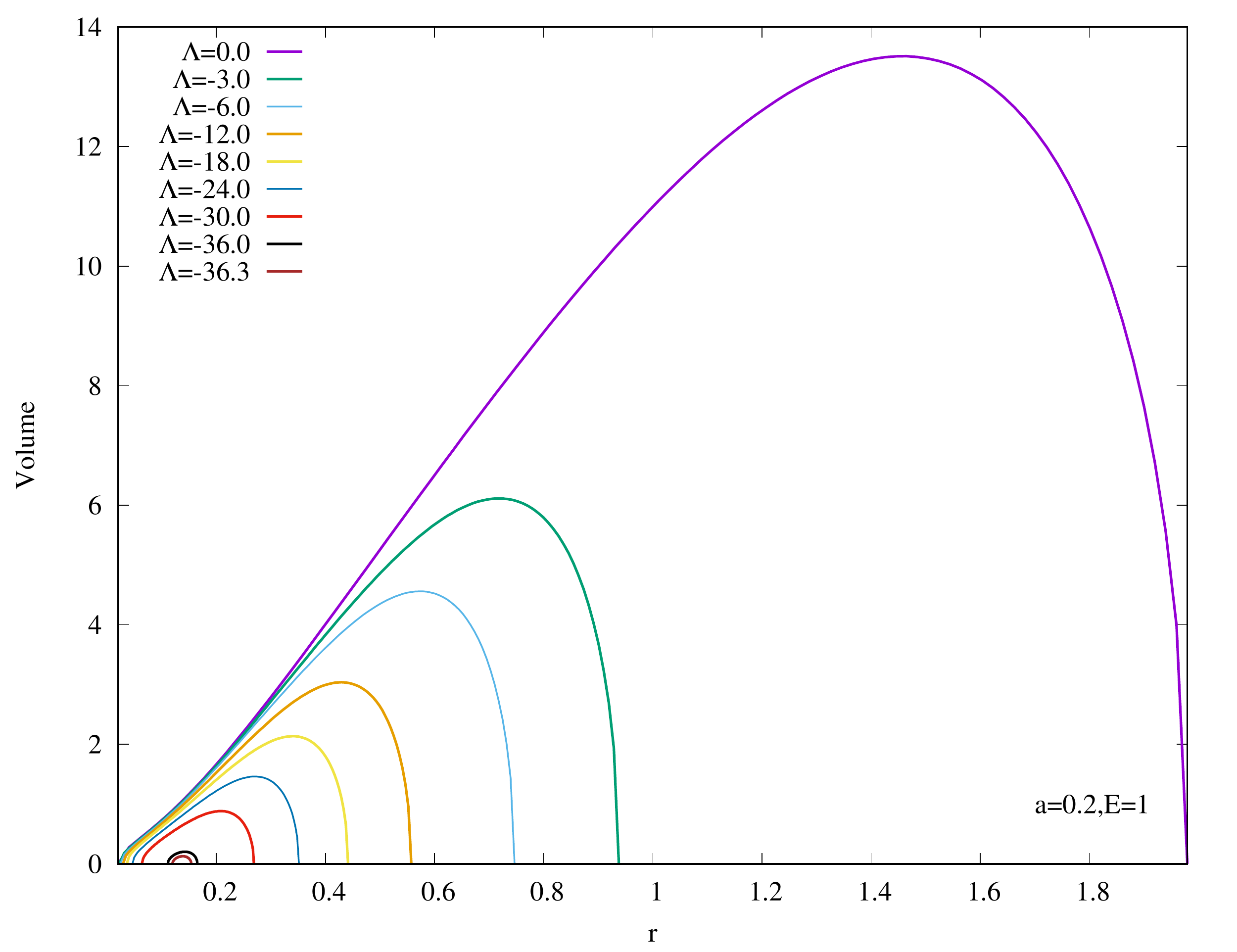}
(b) \includegraphics[angle =270,scale=0.31]{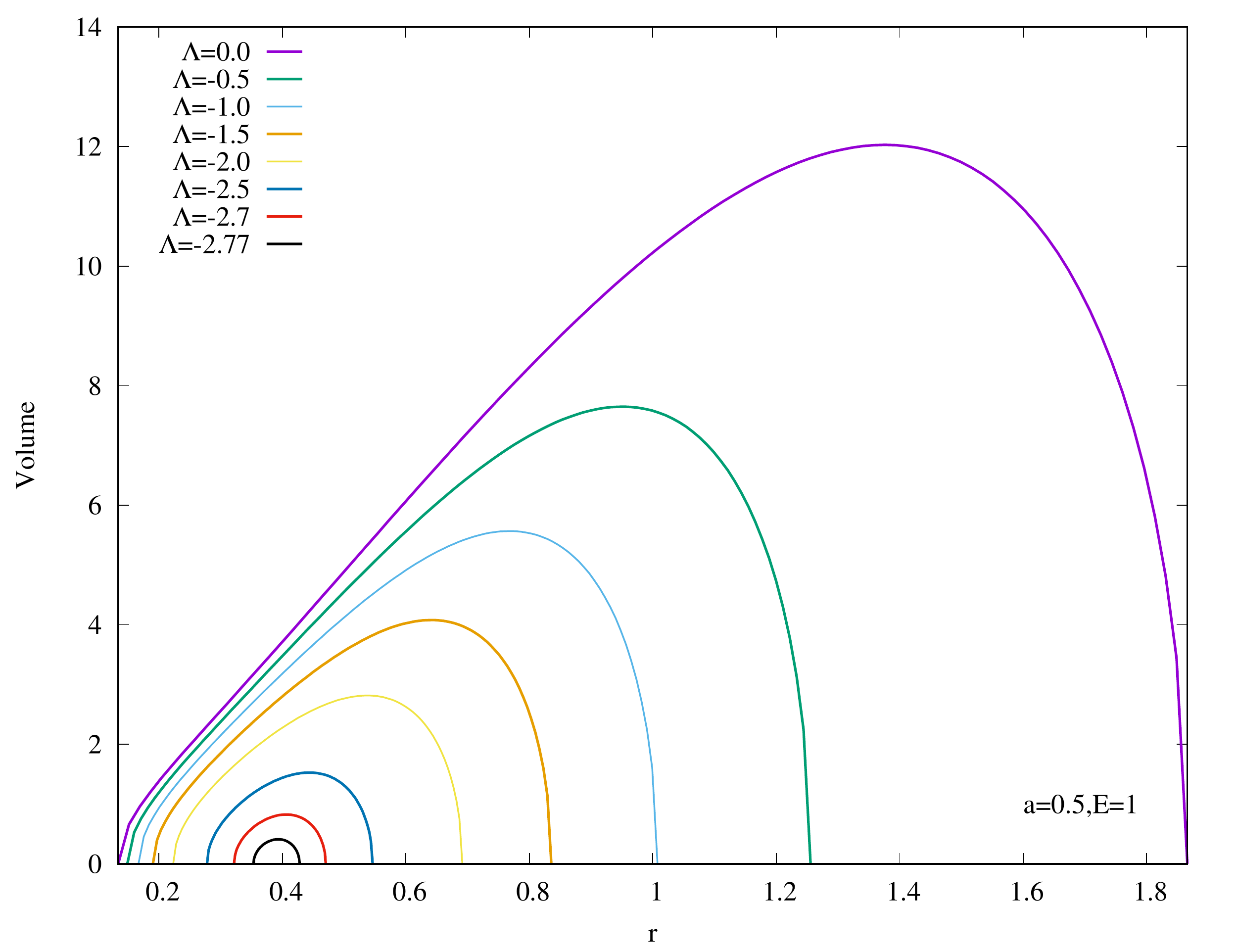} 
}
\mbox{
(c)
\includegraphics[angle =270,scale=0.31]{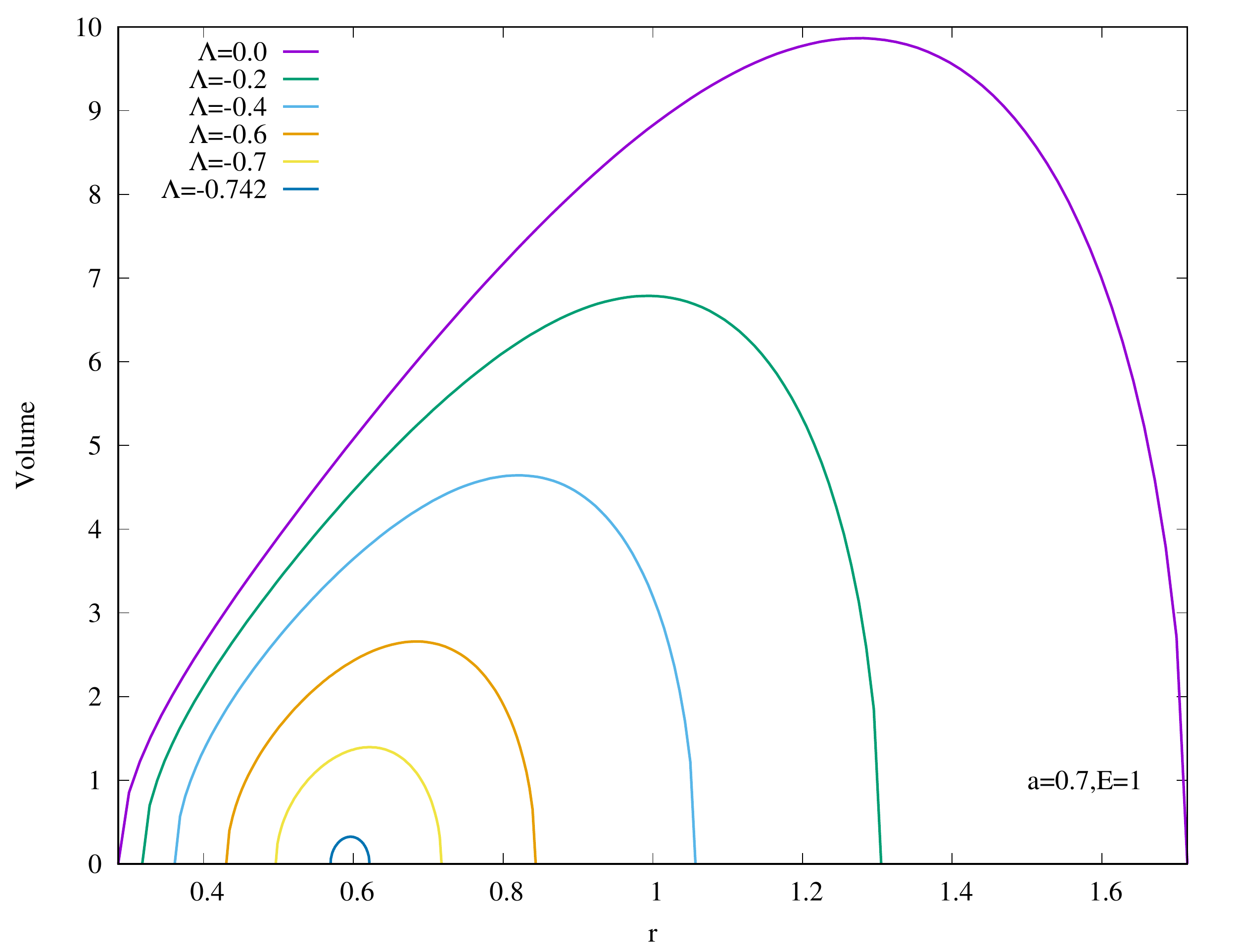}
(d)
\includegraphics[angle =270,scale=0.31]{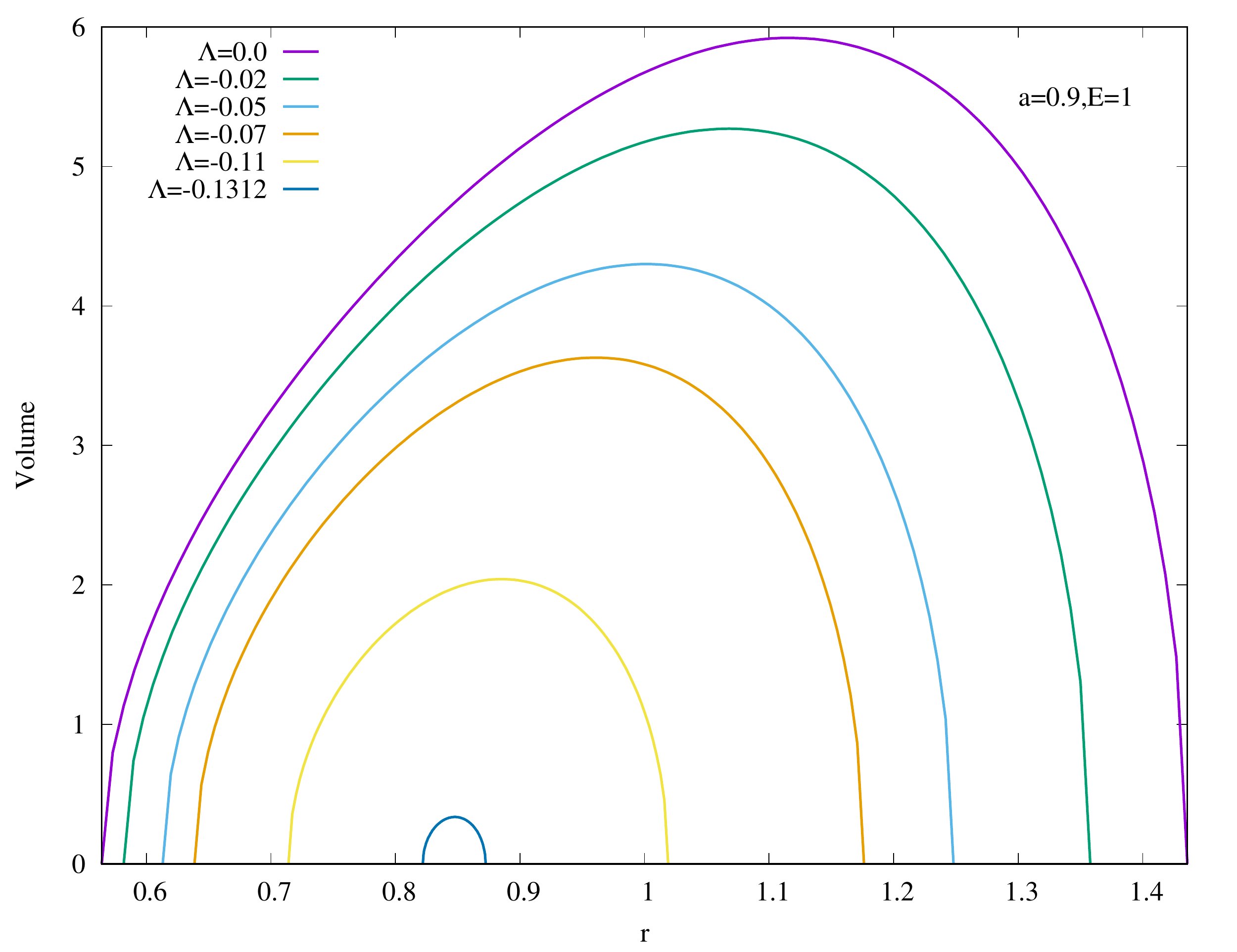}
}
\mbox{
(e)
\includegraphics[angle =270,scale=0.31]{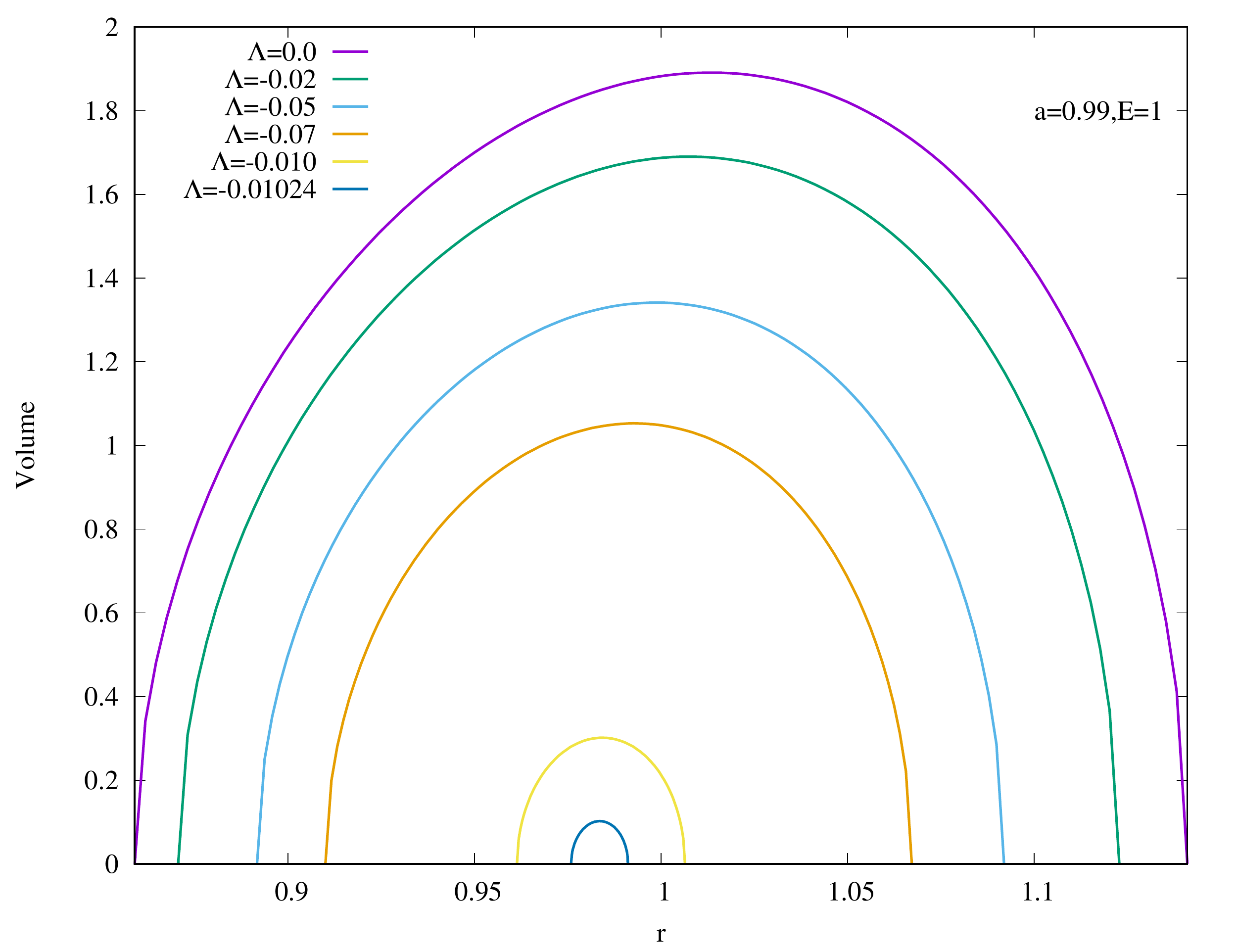}
}
\caption{The interior volume of Kerr--AdS black hole with $E=1$ for (a) $a=0.2$, (b) $a=0.5$, (c) $a=0.7$, (d) $a=0.9$ and (e) $a=0.99$.  \label{Fig5}}
\end{figure}

This is \emph{not} yet the Christodoulou-Rovelli volume since that requires maximizing the expression.
This is neither analytically realistic nor illuminating, therefore we shall follow the approach of \cite{Bengtsson:2015zda} and plot Eq.(\ref{Volume_eqn}), its \emph{peak} then corresponds to the Christodoulou-Rovelli volume.
The volume of Kerr--AdS black hole is shown in Figure \ref{Fig5} for several values of $a$ with $E=1$. 
For all values of $a$, we can see that asymptotically flat Kerr black hole ($\Lambda=0$) has the largest volume.
The volume starts to shrink when $\Lambda$ decreases for fixed $a$, and eventually becomes zero when the black holes become extremal\footnote{This does \emph{not} mean that extremal black holes have zero volume. The reason for non-extremal black hole to have large growing volume is because $r$ coordinate becomes timelike direction inside the black hole. For extremal black holes this does not happen, and so the volume is the ``usual'' volume, much like a sphere in Euclidean space.}. The peak of the volume tends to smaller values of the radial coordinate when  $\Lambda$ decreases for fixed $a$. In addition, the curves do not intersect. If $\Lambda_i < \Lambda_j$, then the curve for $\Lambda_j$ encloses that of $\Lambda_i$.

\begin{figure}[h!]
\centering
\mbox{ 
(a)  \includegraphics[angle =270,scale=0.31]{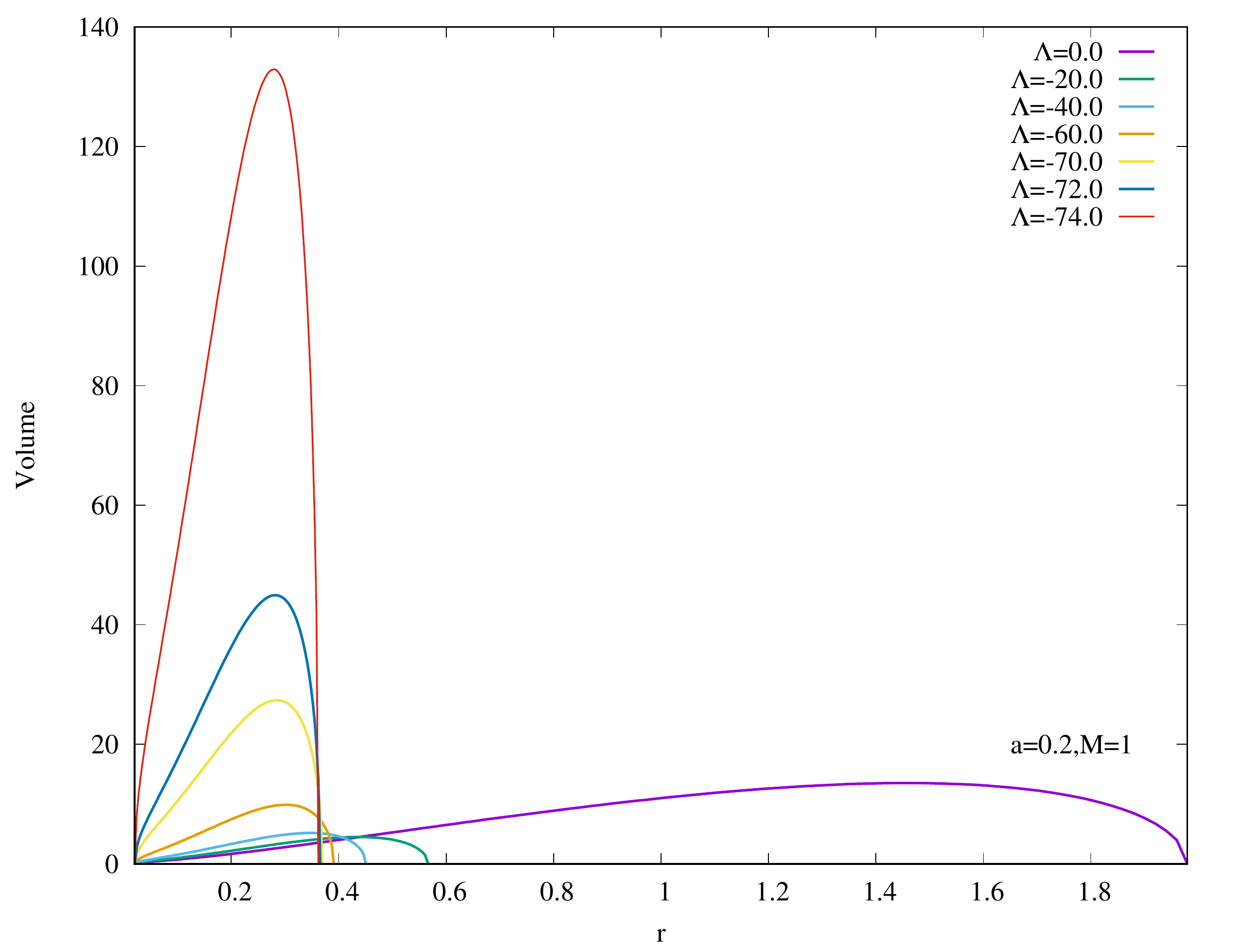}
(b) \includegraphics[angle =270,scale=0.31]{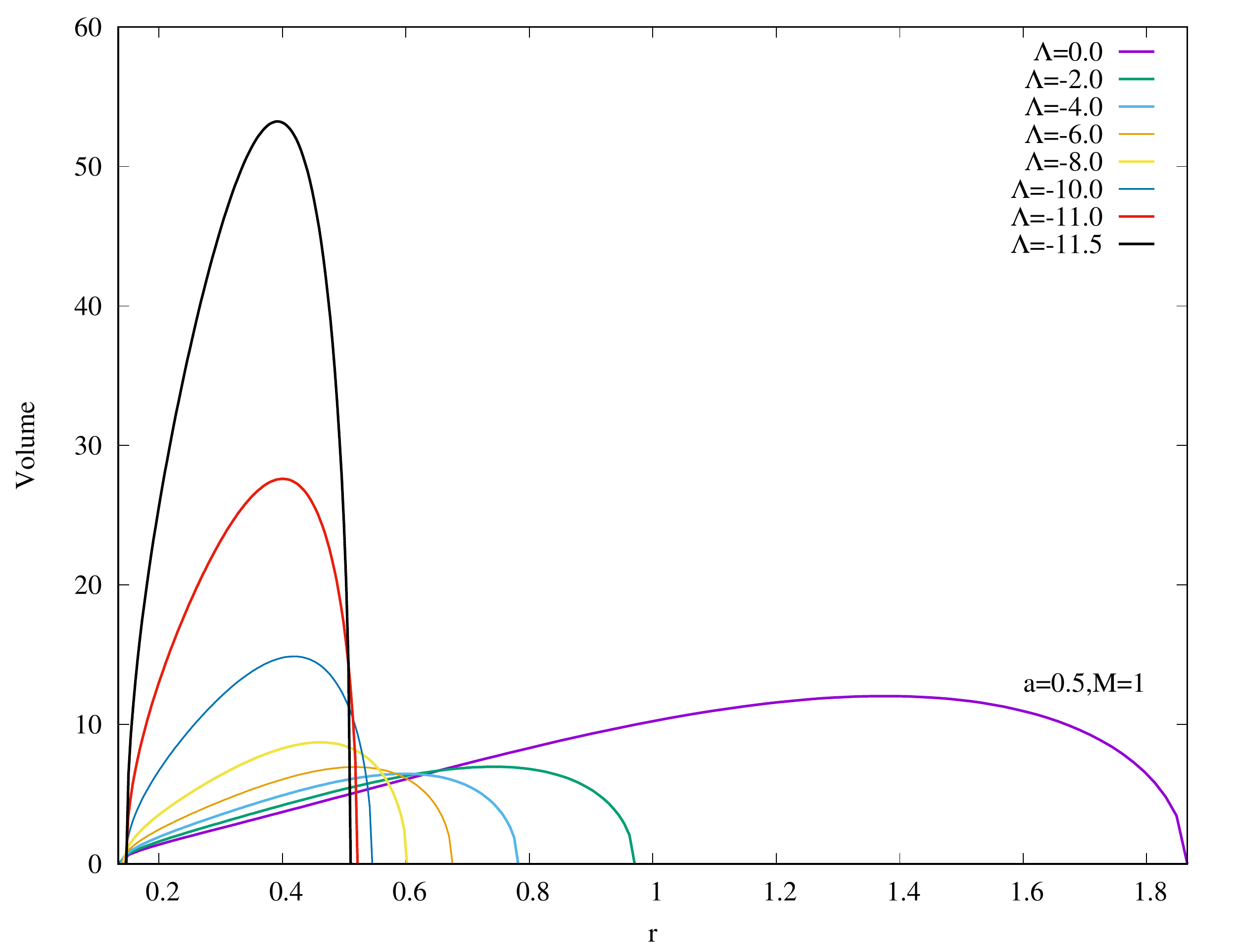} 
}
\mbox{
(c)
\includegraphics[angle =270,scale=0.31]{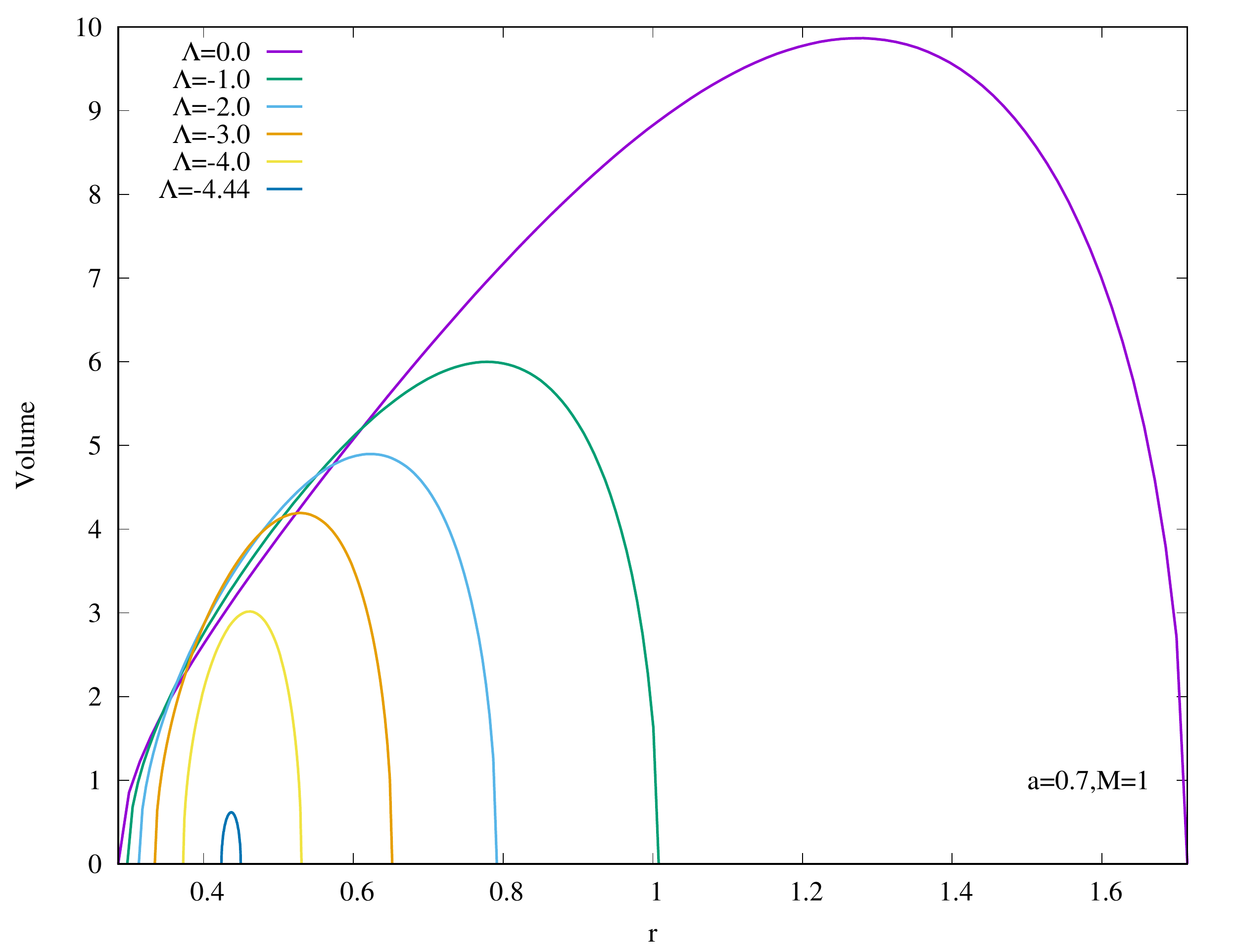}
(d)
\includegraphics[angle =270,scale=0.31]{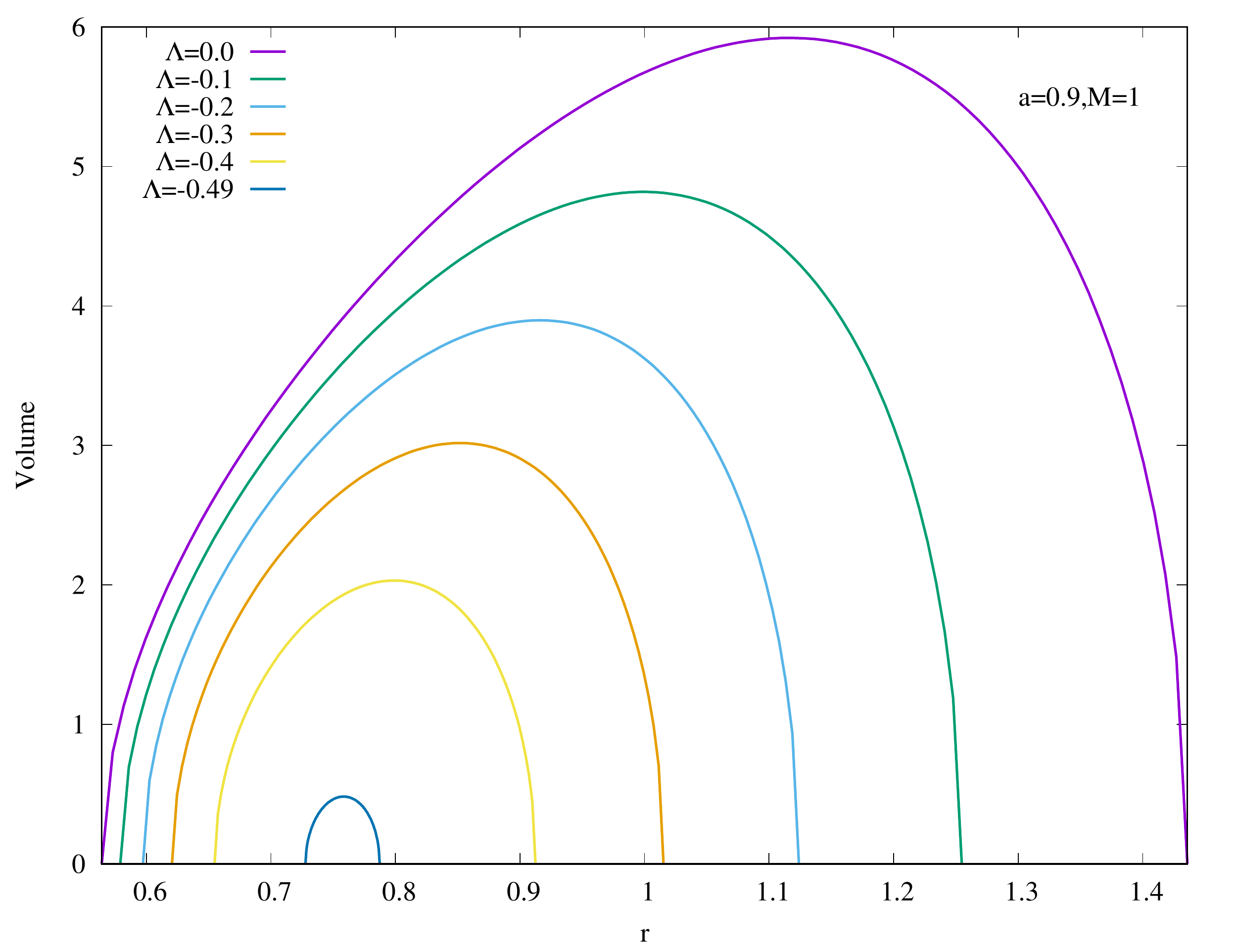}
}
\mbox{
(e)
\includegraphics[angle =270,scale=0.31]{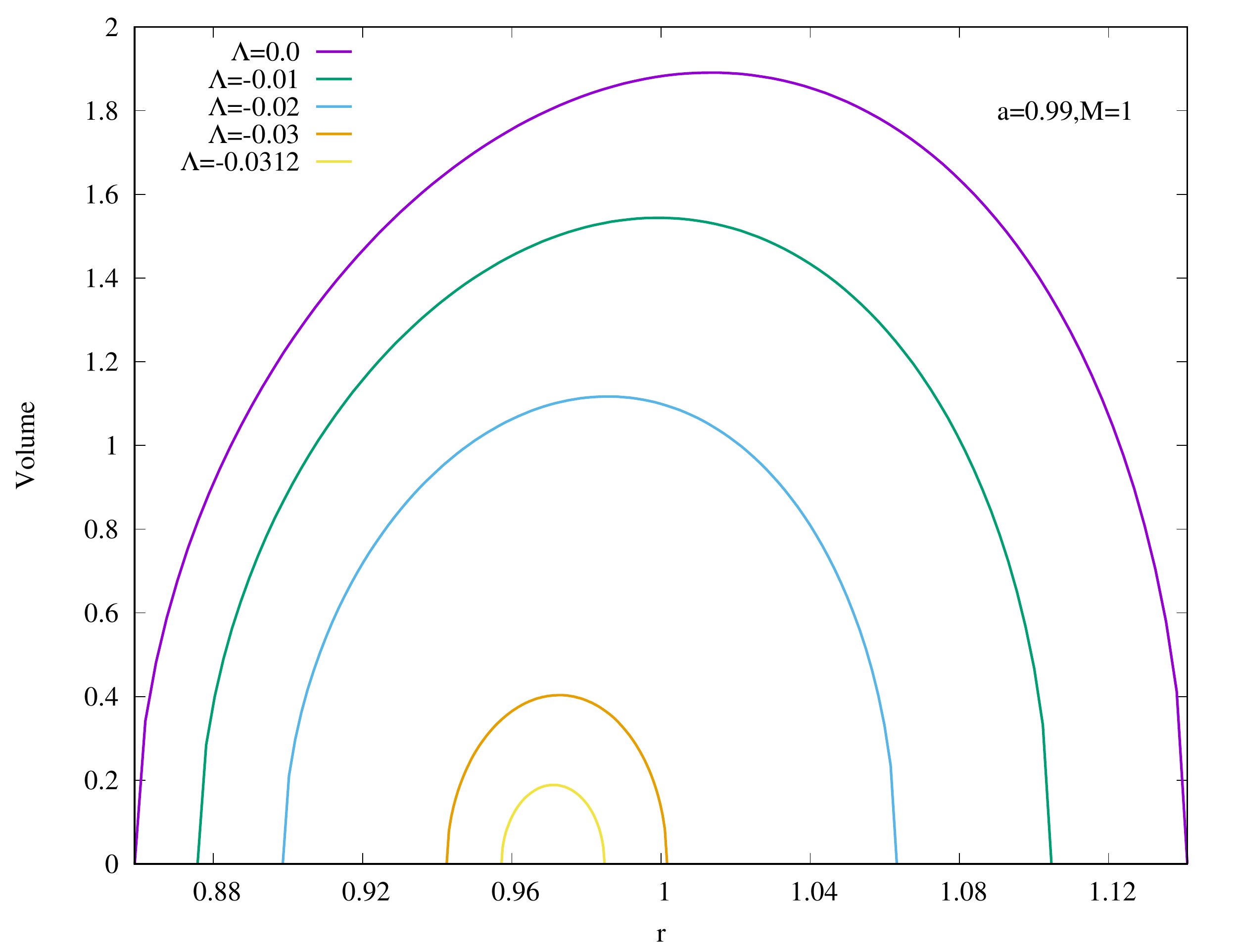}
}
\caption{The interior volume of Kerr--AdS black hole with $M=1$ for (a) $a=0.2$, (b) $a=0.5$, (c) $a=0.7$, (d) $a=0.9$ and (e) $a=0.99$. For black holes with $a \lessapprox  0.64952$, namely (a) and (b), we observe that there are ``double-lobe''. \label{Fig6}}
\end{figure}

As shown in Figure \ref{Fig6}, the rapidly rotating Kerr--AdS black holes with $M=1$ exhibit similar qualitative behavior for the interior volume as the case of $E=1$ if $a$ is large. However, for the slowly rotating ($a \lessapprox  0.64952$) Kerr--AdS black holes, the interior volume exhibits some interesting features as $\Lambda$ is varied. The curves form a ``double lobe'' structure:  starting with $\Lambda=0$ and decreasing $\Lambda$, the curve moves towards smaller value of $r$, shortening in length, and subsequently start to grow in height as it enters the other lobe. For example, for $a=0.5$, the transition happens at around $\Lambda=-8$. This feature is due to the gap between $r_+$ and $r_-$ for sufficiently slow rotating fixed $M$ black holes.
The curves in the case of $M=1$ also typically intersect each other except for  large values of $a$.
The interior volume diverges when $\Lambda$ approaches the bound $a<L$. 

We can also express the interior volume of Kerr--AdS black hole in the unit of $L=1$ \footnote{See footnote 2.}. As shown in the Appendices, the interior volume only exhibit a single lobe structure. For a fixed $a$, the extremal Kerr--AdS black hole possesses the lowest mass $M$ (and also the physical mass $E$) and has almost zero gap between outer and inner horizons, thus they have the smallest interior volume. As $M$ (as well as $E$) increases, the interior volume of Kerr--AdS also increases, since the gap increase. Note that the gap always closes.





\begin{figure}[h!]
\centering
\includegraphics[angle=270,scale=0.31]{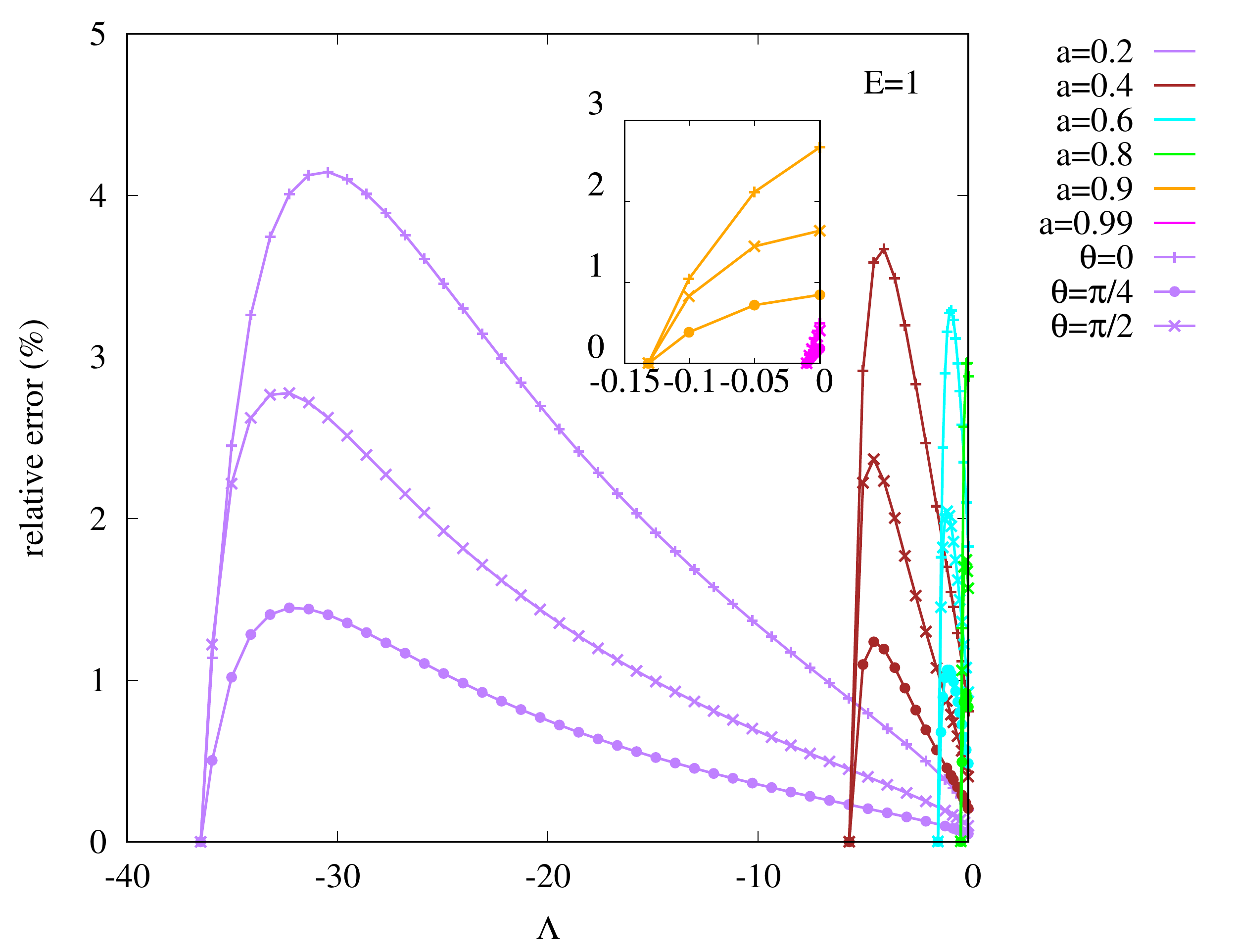} 
\caption{The relative error for the approximation of the location of the peak for Kerr--AdS black hole volume function with $E=1$, as per Eq.~\eqref{volume_approx}.  \label{Fig8}}
\end{figure}

Another method to approximate the peak of the volume function $\text{Vol.}(r)$ can be found in \cite{Couch:2018phr}, in which the authors compute the maximum of the integral, $r_{\text{peak}}$, for arbitrary angle $\theta$ via (their case is for asymptotically flat Kerr, so $\Xi\equiv1$)
\begin{equation} \label{volume_approx}
\frac{\partial }{\partial r} \left( \frac{ \sqrt{ |\Delta_r| \Sigma}}{\Xi} \right) \Bigg|_{\theta=\text{fixed}} = 0 \,.
\end{equation}
The hypersurface obtained in this manner has special mathematical property: it is in some sense a maximal slice with minimal distortion. The details are described in \cite{duncan}.
From Figure \ref{Fig8}, we see that the peak of the interior volume obtained by this approach is very close to the one obtained by the first method regardless of the value of $\theta$: the relative error is less than $5$ percent for the case $E=1$.

\section{Holographic Complexity}

As we have briefly mentioned in Sec.1, the notion of interior volume is not only interesting for its own sake, but also has applications in the holographic setting, specifically in the so-called ``complexity-volume'' conjecture: namely that interior volume is dual to the quantum complexity of the dual field system. It has been argued that there are some ambiguities regarding how this duality is supposed to be defined, since for large black holes (i.e. large compared to the AdS curvature radius $L$), complexity seems to behave as
\begin{equation}
C \sim \frac{V}{\hbar G L}.
\end{equation}
However, for small black holes with horizon $r_+$, complexity goes like
\begin{equation}
C \sim \frac{V}{\hbar G r_+}. 
\end{equation}
Such a seemingly ad hoc rule is one of the reasons why, subsequently, the ``complexity-action'' conjecture was proposed, in which ``action'' refers to the Hilbert-Einstein action restricted to a certain ``Wheeler-DeWitt patch''. However, as pointed out in \cite{Couch:2018phr}, the two seemingly different relations above can be understood rather naturally simply as
\begin{equation}
C \sim \frac{V}{\hbar G \tau_f},
\end{equation}
where $\tau_f$ is the maximal time to fall from the horizon to the (final) maximal slice. Indeed, the infalling time scales like $L$ for large black holes but scales like $r_+$ for small black holes.
Specifically, the infalling time is \cite{Couch:2018phr}, with $\d t = \d \theta = \d \phi =0$, given by
\begin{equation}
 \tau_f = \int_{r_f}^{r_{+}}  \sqrt{\frac{r^2+a^2}{|\Delta_r|}} \d r\,.
\end{equation} 
We will assume that the maximum value is evaluated at $\theta=0$, with the volume formula given by Eq.(\ref{volume_approx}).

\begin{figure} [h!]
\begin{center}
\includegraphics[angle=270,scale=0.31]{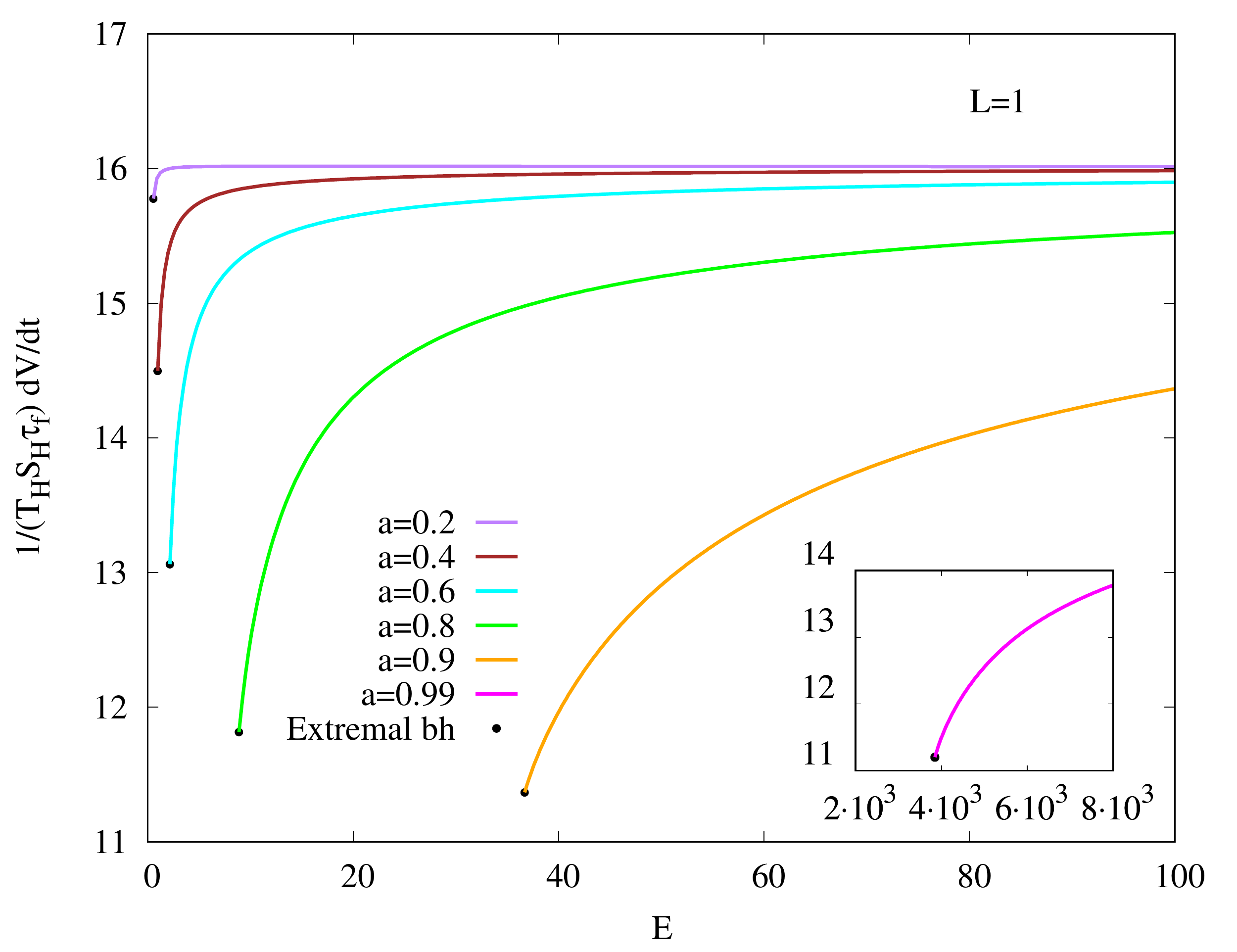} 
\end{center}
\caption{The growth rate of the complexity (as volume) for Kerr--AdS in the unit of $L$. The black dot represents the extremal Kerr--AdS. \label{Fig9}}
\end{figure}

As we can see in Fig.(\ref{Fig9}), the quantity 
\begin{equation}
\mathcal{Q}:=\frac{1}{T_HS_H \tau_f} \frac{\d V}{\d t}
\end{equation}
for any fixed angular momentum parameter does not vary too much as the physical mass $E$ is varied.
Alternatively, for a fixed $E$, increasing $a$ would lower the value of $\mathcal{Q}$ but the value is of the same order of magnitude. This behavior is similar to the asymptotically flat case discussed in \cite{Couch:2018phr}. See Fig.(11) therein. That is, the growth rate of the black hole volume scales approximately as $T_HS_H \tau_f$. So this matches the complexity growth rate of the entropy $\sim T_HS_H$ \cite{1402.5674} if one identifies complexity with $V/G\hbar\tau_f$.

\section{Conclusion}

Rotating black holes in anti-de Sitter spacetimes play important roles in holography (see, e.g. \cite{1305.3267, 1803.02528}). Motivated by the complexity-volume conjecture in holography, as well as the recent interests in black hole chemistry, which treats the cosmological constant as a thermodynamic pressure that is dual to the number of colors in the boundary field theory, we take a closer look at the interior volume of Kerr-AdS black holes and the effect of varying the cosmological constant on the volume (though the actual implications for holography is beyond the scope of the current work). 

While varying the cosmological constant, we have either fixed the physical mass to $E=1$ or the ``geometric mass'' to $M=1$ for simplicity and convenience. Our results remain qualitatively similar for other fixed values of $E$ and $M$.
The physical mass is the mass that enters the laws of thermodynamics, while the ``geometric mass'' gives a better measure for the underlying spacetime geometry. That is, they are good for different purposes. We found that setting $E$ fixed gives qualitatively the same features for the interior volume, compared to the asymptotically flat case. The curves become more symmetric as the rotation parameter $a$ increases. Setting $M$ fixed, however, gives rise to a ``double lobe'' feature in the plot of volume as function of the coordinate radius $r$. Recall that the peak of any given curve in the $\text{Vol}.(r)$ plot corresponds to the Christodoulou-Rovelli volume, the physical interpretation is that the volume decreases monotonically as we decreases $\Lambda$ while holding $M$ fixed. On the other hand, holding $E$ fixed allows the volume to first decrease and then increase as $\Lambda$ is decreased, provided the rotation parameter $a$ is sufficiently small. 

Using the notion of volume discussed in this work, we also evaluated the holographic complexity according to the ``complexity-volume'' conjecture as interpreted in \cite{Couch:2018phr}. The result is that the growth rate of volume is indeed proportional to $\tau_f T_H S_H$, where $\tau_f$ is the infalling time from the event horizon to the ``final'' maximal volume hypersurface. This agrees with the expectation that the complexity in the dual field theory should go like $T_H S_H$ if volume is indeed holographically dual to quantum complexity via $C \sim V/G\hbar \tau_f$.

\section*{Acknowledgement}
XYC would like to thank the hospitaltiy of Center for Gravitation and Cosmology of Yangzhou University during his visit to attend COSPA 2018. XYC would also like to thank Dong-han Yeom for helping him adapting to life in Busan. He thanks the National Research Foundation of Korea (grant no. 2018R1D1A1B07049126) for funding. YCO thanks the National Natural Science Foundation of China (No.11705162, No.11922508) and the Natural Science Foundation of Jiangsu Province (No.BK20170479) for funding support. He also thanks Ingemar Bengtsson for related discussions.

\newpage

\appendix

\section{ Interior of Kerr--AdS with $M$ (in the unit of $L$)}

\begin{figure}[h!]
\centering
\mbox{ 
(a)  \includegraphics[angle =270,scale=0.31]{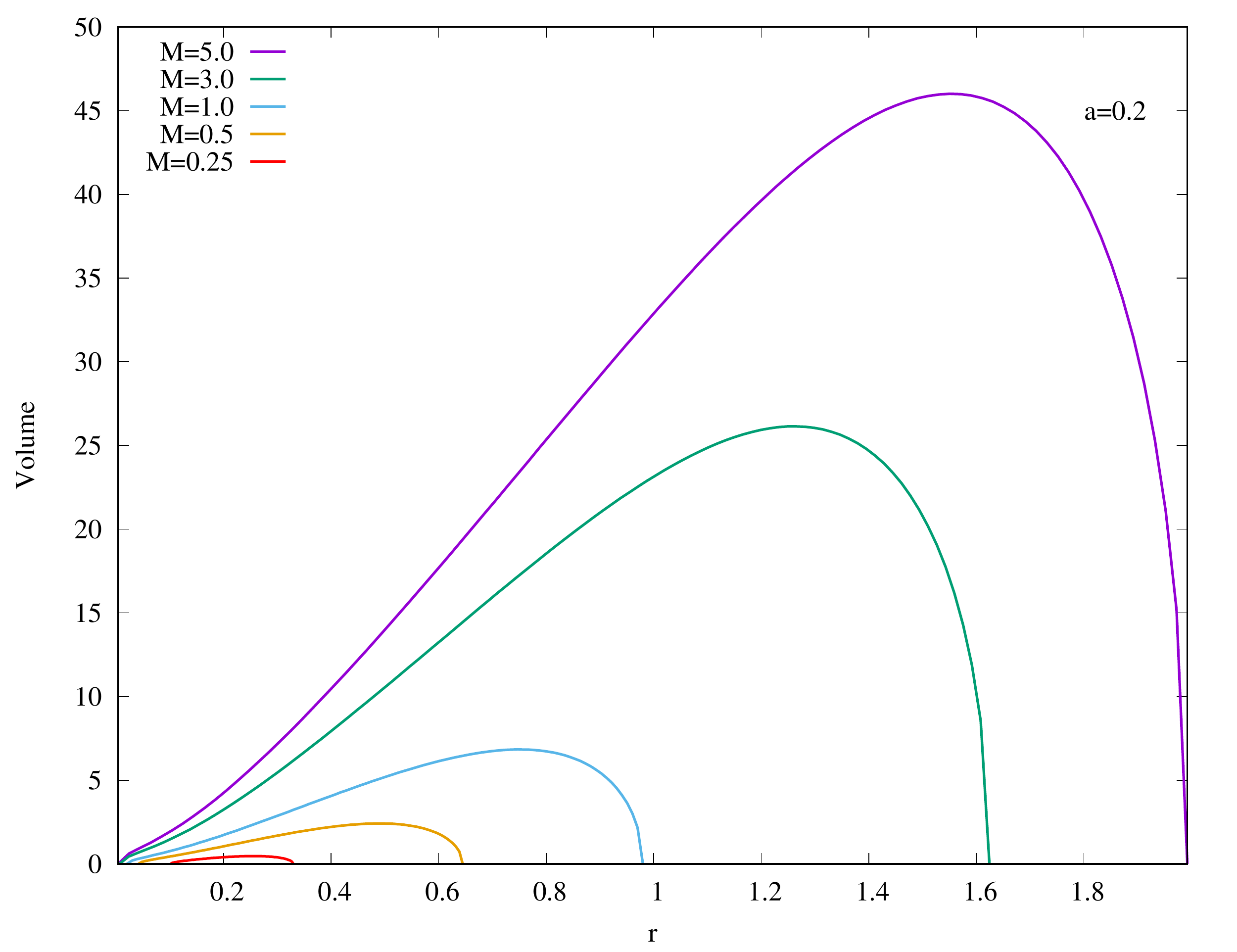}
(b) \includegraphics[angle =270,scale=0.31]{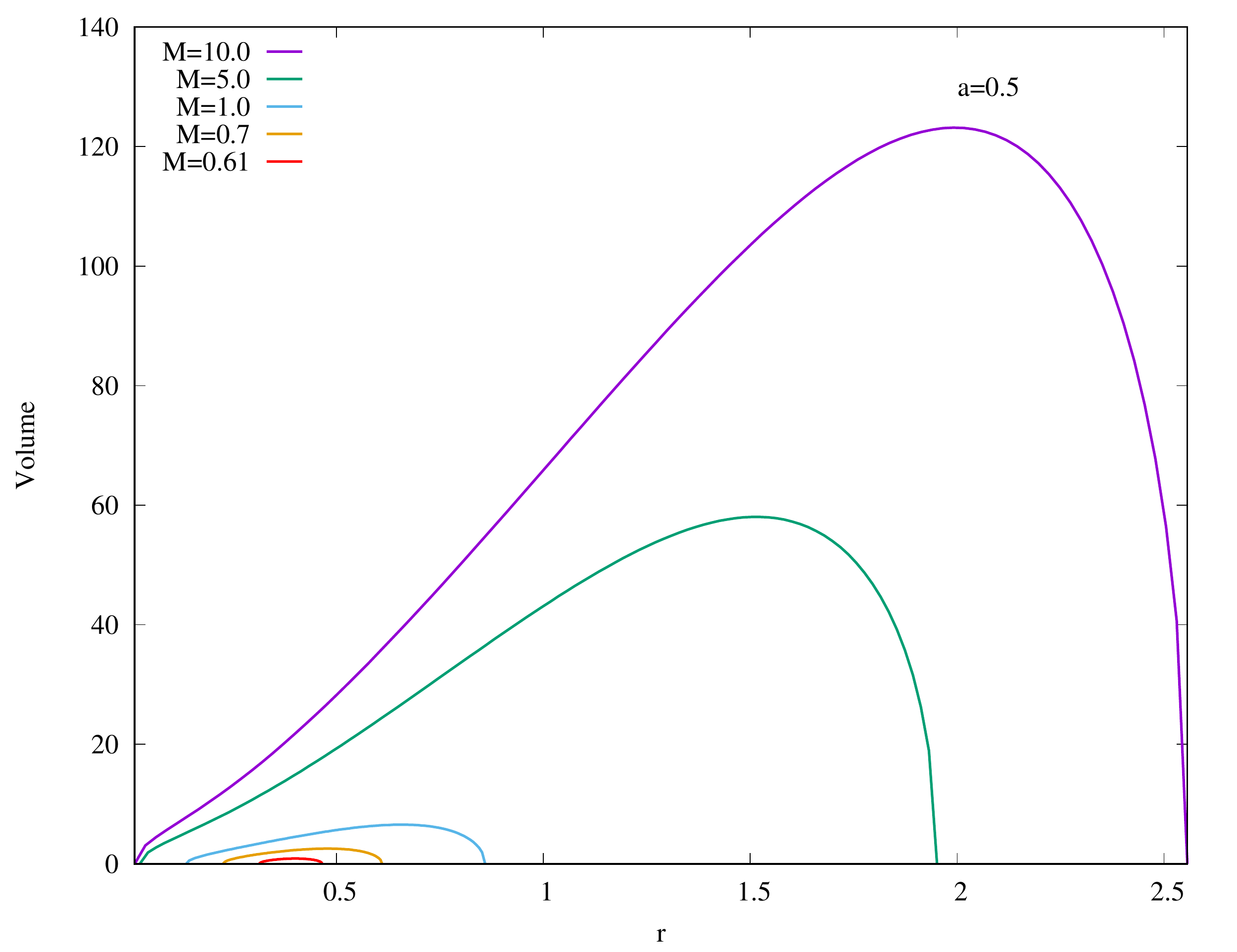} 
}
\mbox{
(c)
\includegraphics[angle =270,scale=0.31]{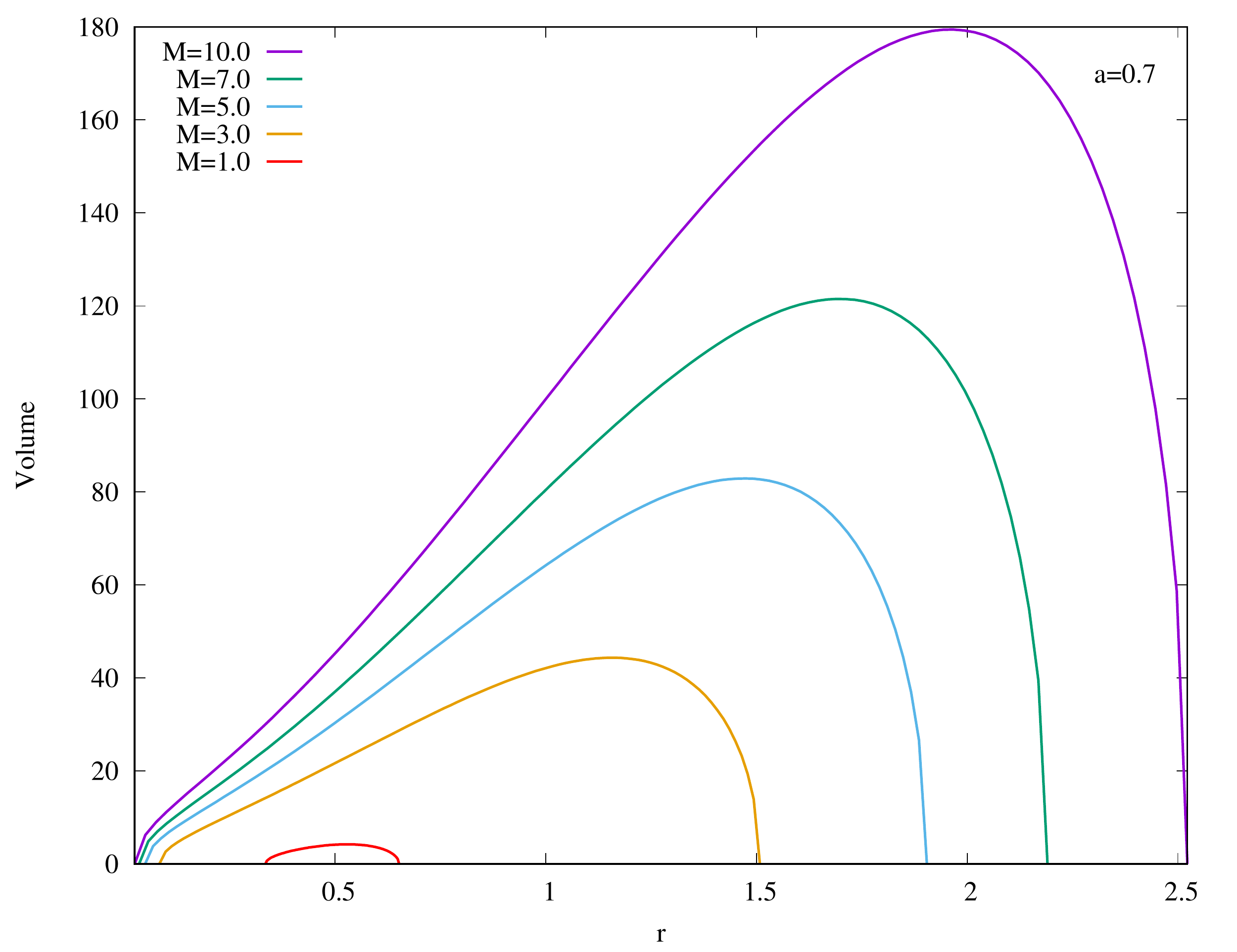}
(d)
\includegraphics[angle =270,scale=0.31]{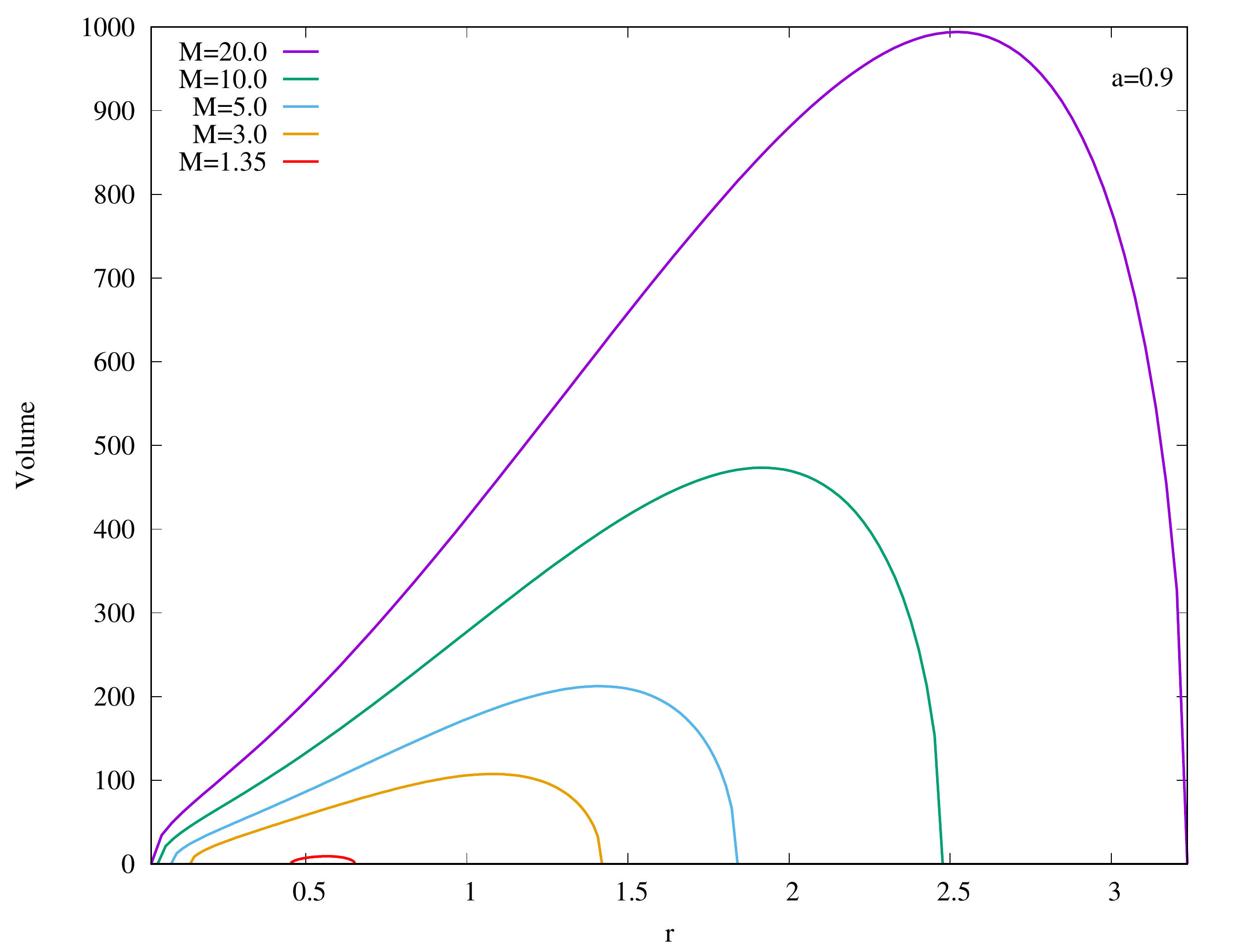}
}
\mbox{
(e)
\includegraphics[angle =270,scale=0.31]{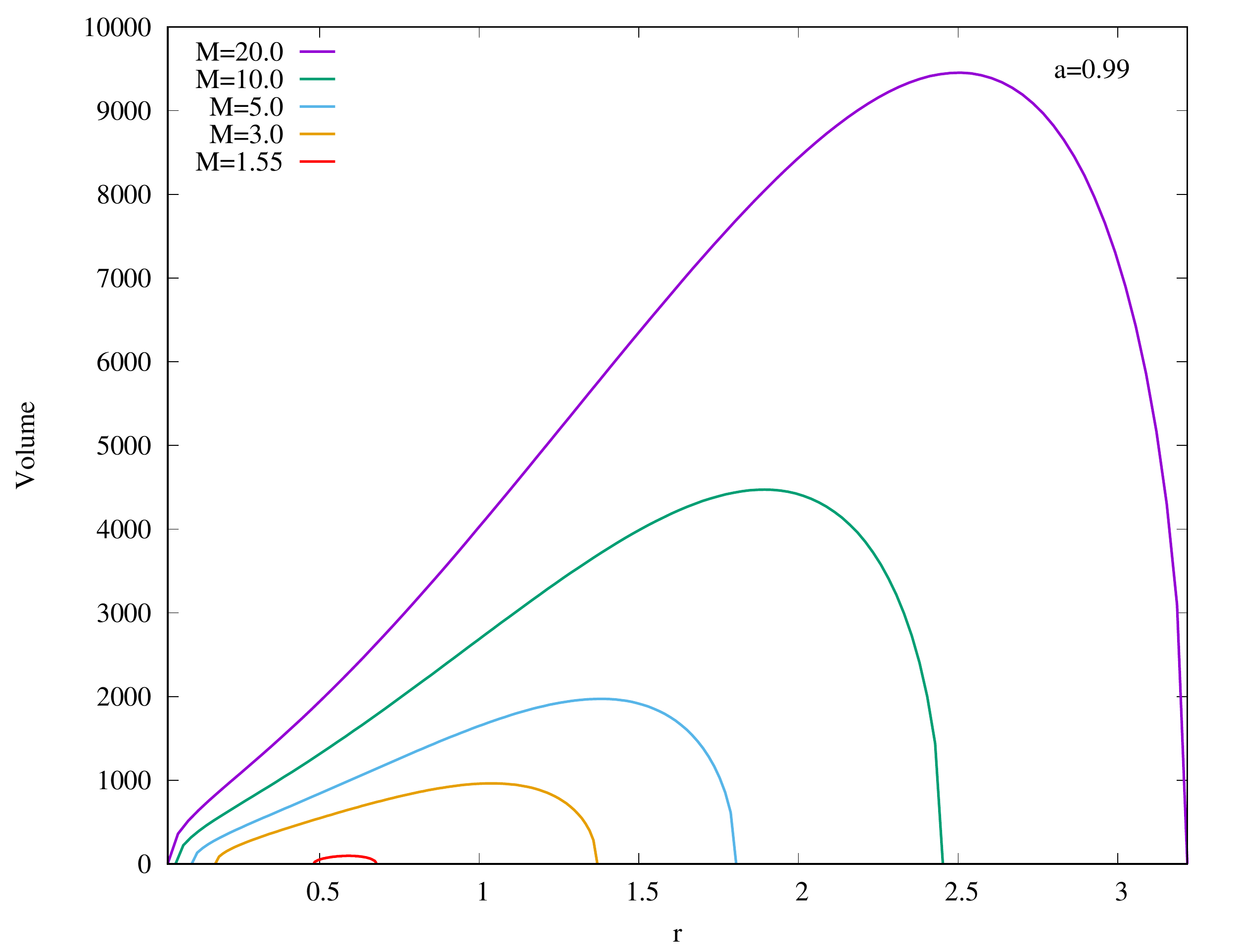}
}
\caption{The interior volume of Kerr--AdS black hole with the mass $M$ in the unit of $L$ for (a) $a=0.2$, (b) $a=0.5$, (c) $a=0.7$, (d) $a=0.9$ and (e) $a=0.99$.  \label{Fig10}}
\end{figure}

\newpage
\section{ Interior of Kerr--AdS with $E$ (in the unit of $L$)}

\begin{figure}[h!]
\centering
\mbox{ 
(a)  \includegraphics[angle =270,scale=0.31]{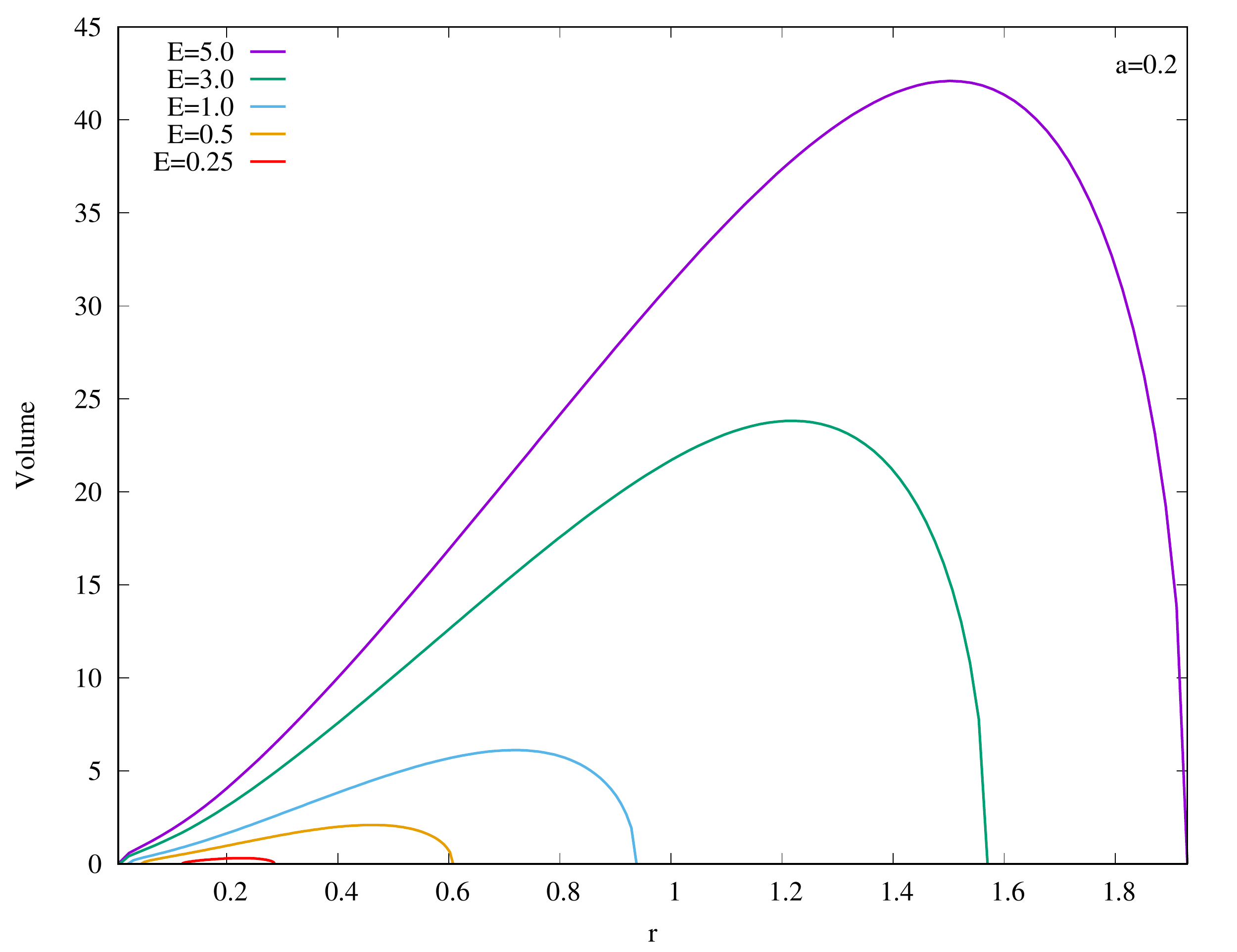}
(b) \includegraphics[angle =270,scale=0.31]{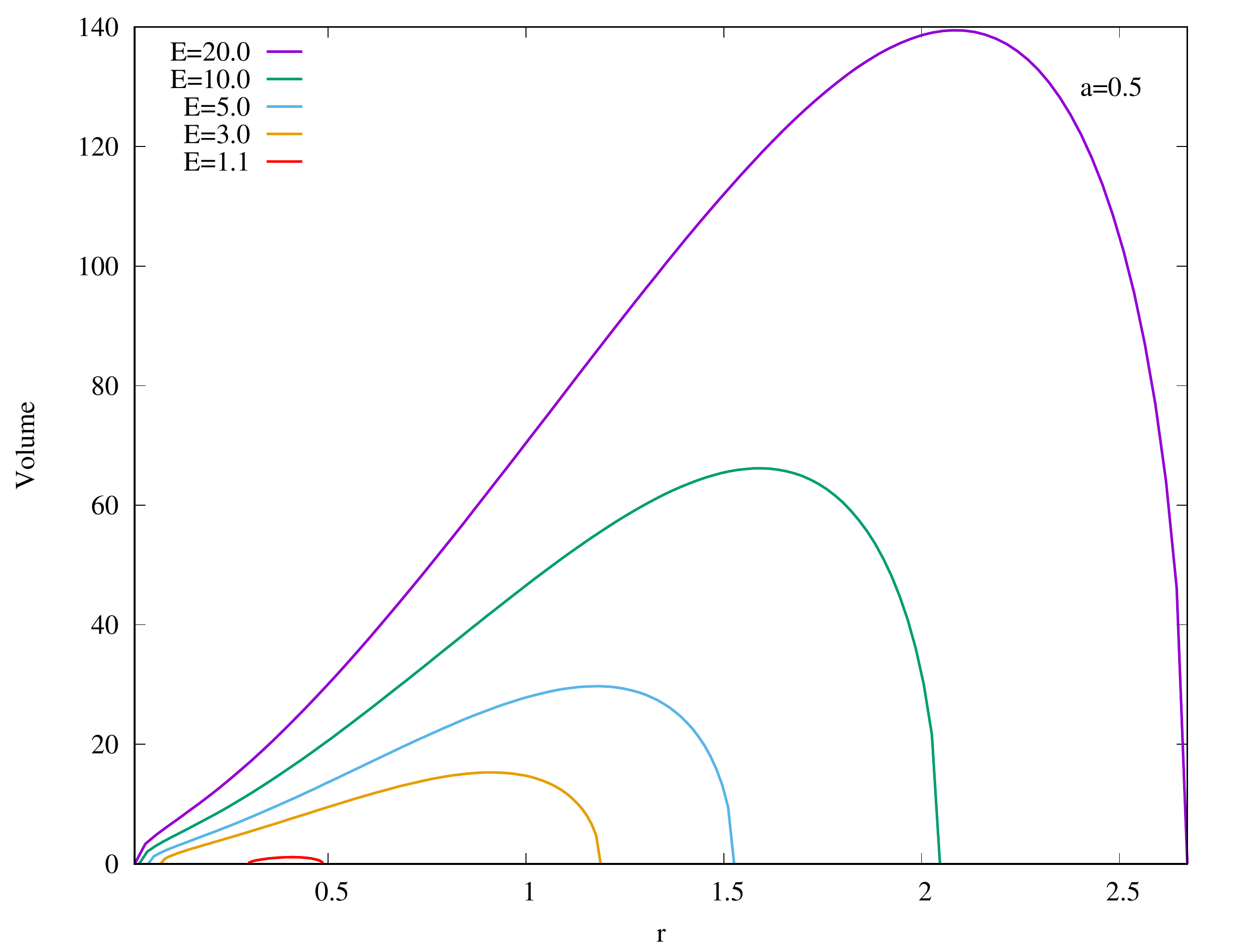} 
}
\mbox{
(c)
\includegraphics[angle =270,scale=0.31]{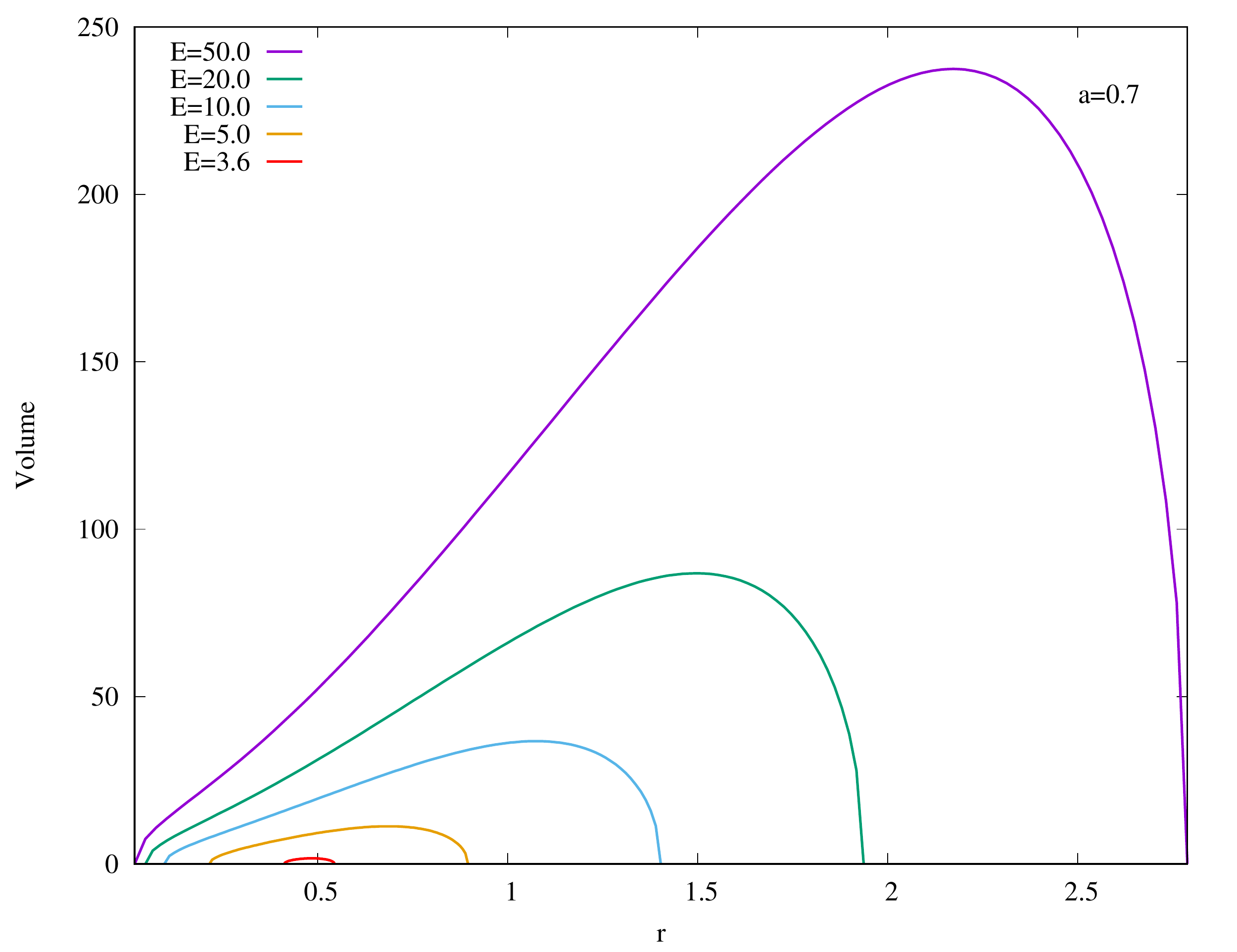}
(d)
\includegraphics[angle =270,scale=0.31]{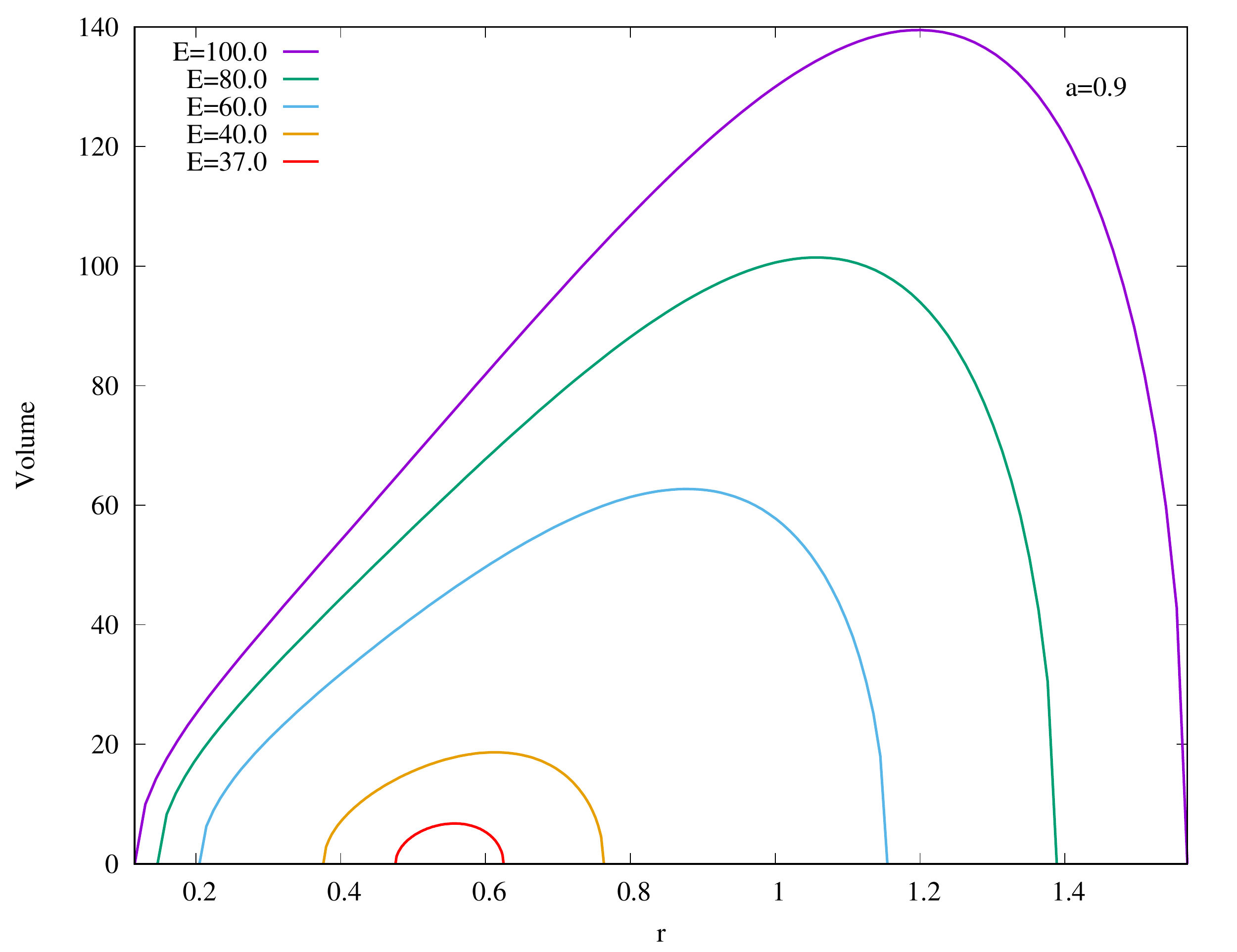}
}
\mbox{
(e)
\includegraphics[angle =270,scale=0.31]{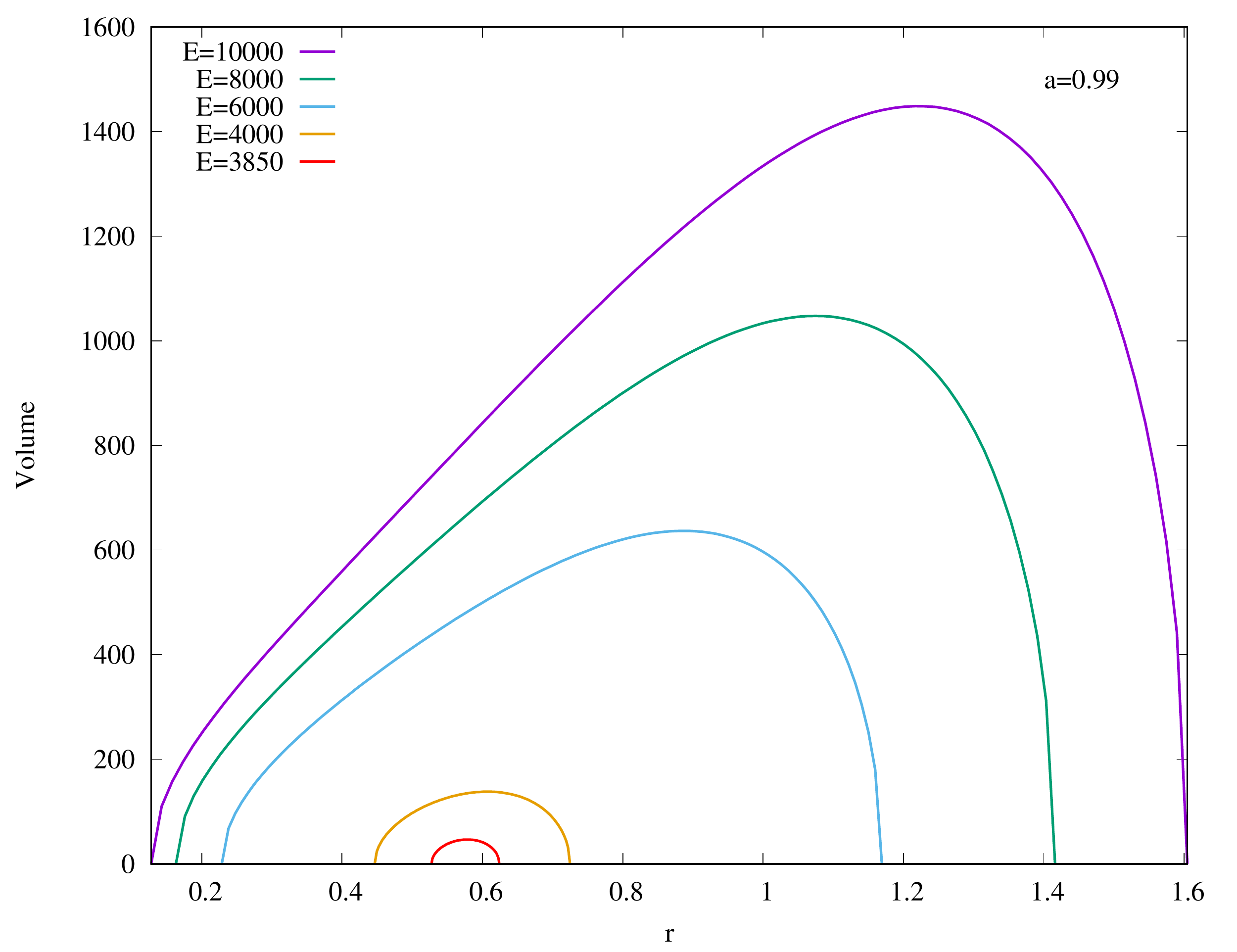}
}
\caption{The interior volume of Kerr--AdS black hole with the physical mass $E$ in the unit of $L$ for (a) $a=0.2$, (b) $a=0.5$, (c) $a=0.7$, (d) $a=0.9$ and (e) $a=0.99$.  \label{Fig11}}
\end{figure}

\end{document}